\documentclass{aa}
\usepackage{booktabs,caption,fixltx2e}
\usepackage[flushleft]{threeparttable}
\usepackage{graphicx}
\usepackage{natbib}

\begin{document}

\title{Age -- Metallicity relation in the MCs clusters \footnote{Based on observations made with the Danish 1.54 m and ESO 3.6m Telescopes
at La Silla Observatory, Chile.}
}

\author{E. Livanou\inst{1}
\and A. Dapergolas\inst{2}
\and M. Kontizas\inst{1}
\and B. Nordstr\"{o}m\inst{3}
\and E. Kontizas\inst{2}
\and J. Andersen\inst{3,5}
\and B. Dirsch\inst{4}
\and A. Karampelas\inst{1}}

\offprints{E. Livanou \\
\email{elivanou@phys.uoa.gr}}


\institute{Section of Astrophysics Astronomy \& Mechanics,
Department of Physics, University of Athens, GR-157 83 Athens, Greece
\and
Institute of Astronomy and Astrophysics, National Observatory of Athens, P.O.
Box 20048, GR-118 10, Greece
\and
Niels Bohr Institute Copenhagen University, Astronomical Observatory, Juliane Maries Vej 30, DK 2100 Copenhagen
\and
Facultad de Ciencias Astronomicas y Geofisicas, Universidad Nacional de La Plata, Argentina
\and
Nordic Optical Telescope, Apartado 474, E-38700 Santa Cruz de La Palma, Spain}

\date{Received date / accepted}

\abstract
{}
{To investigate a possible dependence between age and metallicity in the Magellanic Clouds (MCs) 
from a study of small open star clusters, using Str\"{o}mgren  photometry. Our goal is to trace 
evidence of an age metallicity relation (AMR) and correlate it with the mutual interactions of 
the two MCs. Our aim is also to correlate the AMR with the spatial distribution of the clusters. 
In the Large Magellanic Cloud (LMC), the majority of the selected clusters are young (up to 1 Gyr) 
and our aim is to search for an AMR at this epoch which has not been much studied.}
{We report on results for 15 LMC and 8 Small Magellanic Cloud (SMC) clusters, scattered all over 
the area of these galaxies, to cover a wide spatial distribution and metallicity range. 
The selected LMC clusters were observed with the 1.54m Danish Telescope in Chile, using the Danish 
Faint Object Spectrograph and Camera (DFOSC) with a single 2k$\times$2k CCD, whereas the SMC clusters 
were observed with the ESO 3.6m Telescope also in Chile with the ESO Faint Object Spectrograph and
Camera (EFOSC). The obtained frames were analysed with the conventional DAOPHOT and IRAF software.
We used Str\"{o}mgren filters in order to achieve reliable metallicities from photometry. Isochrone 
fitting was used in order to determine the ages and metallicities.}
{The AMR for the LMC displays a metallicity gradient, with higher metallicities for the younger ages.
The AMR for LMC-SMC star clusters shows a possible jump in metallicity and a considerable increase at 
about 6$\times$10$^8$yr. It is possible that this is connected to the latest LMC-SMC interaction. 
The AMR for the LMC also displays a metallicity gradient with distance from the centre.
The metallicities in SMC are lower, as expected for a metal poor host galaxy.}
{}

\keywords{galaxies: Magellanic Clouds 
,abundances,clusters}

\maketitle

\section{Introduction}

The age metallicity relation (AMR) is a very important tool for understanding  the chemical 
evolution of a galaxy. The Magellanic Clouds (MCs), our nearest galaxies, offer ideal targets for  such 
studies not only because of their proximity but also due the fact that LMC, SMC and the Milky Way form a 
system of interacting galaxies. It is therefore important to trace the influence of this  
interaction  on the star formation and the chemical evolution. 

The correlation between the age of the stellar clusters in the MCs and their close encounters 
with each other and our Galaxy has already been discussed in previous papers. 
In the LMC a sudden rise in the  star formation rate (SFR) is traced
2 to 4 Gyr ago  (Elson et al. \cite{Elson97}; Geisler et al. \cite{Geisler97}) 
preceded by either a constant lower SFR (Geha et al. \cite{Geha97})  
or possibly a virtual gap as manifested by the cluster age distribution 
(Da Costa \cite{DaCosta91}; van den Bergh \cite {vandenbergh91}).
The dramatic increase in the star formation due to the recent  interaction has been revealed 
in the morphological evolution of the LMC and SMC (Maragoudaki  et al. \cite{Fotini98}, \cite{Fotini2001}).

More recently,
Pietrzynski \& Udalski \cite{PU00}, found that the distribution of cluster ages in both galaxies 
revealed a peak at 100 Myr, which may be connected with the last encounter of the LMC and the
SMC.
Chiosi et al. \cite{Chiosi06}, using isochrone fitting for 311 young clusters,
report two enhancements of star formation, between
100$-$150 Myr and between 1 and 1.6 Gyr, and conclude that the last tidal interaction
between the MCs has triggered the formation of both clusters and field 
stars.
Moreover, Glatt et al. \cite{Glatt10}, find two periods of enhanced cluster formation at 125 Myr
and 800 Myr in the LMC and at 160 Myr and 630 Myr in the SMC. The cluster ages were determined by fitting 
Padova and Geneva isochrones.

The gradient in metallicity is providing  information on the chemical 
evolution of the two galaxies. A systematic radial metallicity trend is found in the 
cluster system of the  LMC (Kontizas, et al. \cite{Kontizas93}) from a sample of clusters up 
to 8 Kpc from the centre.
The sudden rise in the SFR could explain 
the corresponding sudden rise in the metallicity possibly connected to a 
former close encounter with the Milky  Way which took place $\sim$1.5 Gyr ago. The 
metallicity and SFR connected to this event  has been observed both from the metallicity in the 
clusters (Olszewski et al. \cite{olszewski96}; Geisler et  al. \cite{Geisler97}) and from 
$\alpha$-particle elements in planetary nebulae (Dopita \cite{Dopita97}).  Considering that a more 
recent encounter occurred 0.2 to 0.4 Gyr ago (Gardiner \& Noguchi \cite{Gardiner96};  Kunkel 
et al. \cite{Kunkel2000}) it is very interesting to see if these two events have left 
traces in the AMR. Dirsch et al. \cite{Boris2000} have determined the metallicity of six 
LMC populous clusters and their  fields from Str\"{o}mgren photometry.
They propose that their AMR predicts a less steep increase  in the metallicity in earlier 
time than that found by Pagel \& Tautvaisiene \cite{Pagel99}.
Piatti  \& Geisler \cite{Piatti12},
present age and metallicity estimates of 5.5 million stars distributed throughout the LMC.
They find evidence of AMR for the ages up to 1 Gyr, but no significant metallicity gradient between 5 and 12 Gyr.

In the SMC, Da Costa \& Hatzidimitriou \cite{DaCostaHatz98}, determined metallicities from 
spectra of red giants at the Ca II triplet. The resulting AMR is generally consistent with
that for a simple model of chemical evolution, scaled to the present-day SMC mean abundance and gas
mass fraction. Using the same method, Carrera et al. \cite{Carrera08},
trace a metallicity gradient for the first time in the SMC. They also relate a spatial
metallicity gradient to an age gradient, in the sense that more metal-rich stars, which are also younger,
are concentrated in the central regions of the galaxy.
Piatti \cite{Piatti11}
presents age and metallicity estimates of 11 SMC clusters obtained from CCD Washington CT1T2 photometry.
Two enhanced star formation periods are found at 2 Gyr and at 5-6 Gyr, which have taken place throughout 
the entire galaxy. However they notice absence of age metallicity gradient and a relative spread 
in metallicity for clusters older than 7 Gyr.

Therefore it seems worthwhile to investigate the AMR for the MCs, especially for the youngest (up to 1 Gyr) 
clusters in the LMC  and search for traces due to the most recent interactions.
In section 2 of this paper we describe the observational characteristics and data reduction, while in section 3 we 
present the derived ages and metallicities and discuss our results.
Our conclusions are given in section 4.

\section{Observations -- Reductions}

The MCs possess a large population of stellar 
clusters of a whole range of ages. Small open LMC clusters offer homogeneous and ideal 
targets for this investigation. Their  small central density allow us to derive cluster 
parameters with CCD Str\"{o}mgren photometry with small telescopes and reasonable integration times.

Four observing runs at La Silla in Chile were granted to this project. We observed the LMC clusters 
with the 1.54m Danish Telescope, using the Danish Faint Object Spectrograph and Camera (DFOSC) with
a single 2k$\times$2k CCD which was matched
the RCA SIO 501 EX CCO (optimum final pixel size of 0."4).
The full field covered by the instrument is 13.'7$\times$13.'7. 
The ESO 3.6m Telescope was used to observe the SMC with the ESO Faint Object Spectrograph and Camera (EFOSC).
This CCD camera can be used as a very efficient instrument for wideband photometry of crowded stellar fields.
EFOSC size is 1024 x 1024 pixels, with total field of view 5.'4$\times$5.'4 and optimum final pixel size of 0."32.
The observations took place in various intervals between December 1997 and December 2002 (Table~\ref{Obsrerv}).

We used 
the three Str\"{o}mgren filters $y, b, v$ in order to be able to reach as faint as possible and search for 
the oldest small clusters in the LMC periphery. It was neither possible to use the $u$ filter 
nor the $\rm\beta$ filters. The obtained frames have been reduced in the conventional 
way by DAOPHOT from both IRAF and MIDAS packages.

\begin{table*}[h!]
\caption{Observing log. The KMHK clusters are named from Kontizas et al. \cite{Kontizas90} 
whereas KMK clusters are named from Kontizas et al. \cite{Kontizas88}}
\begin{tabular}{lccllll} 
\hline
{$Name$} & {$RA$} & {$DEC$ } & {$exp. time$}  & {$Frames$} & {$Date$} \\
{ } & {h\hspace{.02in} m\hspace{.05in} s} & { d\hspace{.07in} m\hspace{.06in} s } & {y\hspace{.12in} b\hspace{.12in} v}  & {y\hspace{.05in} b\hspace{.05in} v} & {} \\
\hline

KMK1       &5 03 48  &-69 09 44  &15\hspace{.05in} 30\hspace{.05in} 30 &2\hspace{.05in} 2\hspace{.05in} 2 &Dec. 8,9 2002 \\
KMK3       &5 03 45  &-69 05 33  &15\hspace{.05in} 30\hspace{.05in} 30 &2\hspace{.05in} 2\hspace{.05in} 2 &Dec. 8,9 2002 \\
KMK8       &5 04 29  &-69 09 21  &15\hspace{.05in} 30\hspace{.05in} 30 &2\hspace{.05in} 2\hspace{.05in} 2 &Dec. 8,9 2002 \\
KMK32      &5 10 20 & -68 52 45 &15\hspace{.05in} 25\hspace{.05in} 30 &2\hspace{.05in} 2\hspace{.05in} 2 &Dec. 28 1998 \\
HS153      &5 10 30 & -68 52 21 &15\hspace{.05in} 25\hspace{.05in} 30 &2\hspace{.05in} 2\hspace{.05in} 2 &Dec. 28 1998 \\
KMK49      &5 21 10 & -69 56 25 &20\hspace{.05in} 30\hspace{.05in} 30 &2\hspace{.05in} 2\hspace{.05in} 2 &Jan. 3 1999 \\
KMK50      &5 21 23 & -69 54 34 &20\hspace{.05in} 30\hspace{.05in} 30 &2\hspace{.05in} 2\hspace{.05in} 2 &Jan. 3 1999 \\
SL36       &4 46 09 & -74 53 19 &15\hspace{.05in} 30\hspace{.05in} 30 &2\hspace{.05in} 2\hspace{.05in} 2 &Dec. 14,15 1999 \\
SL620      &5 36 29 & -74 24 18 &15\hspace{.05in} 30\hspace{.05in} 30 &2\hspace{.05in} 2\hspace{.05in} 2 &Dec. 16 1999 \\
KMHK81     &4 45 13& -75 07 00 &15\hspace{.05in} 30\hspace{.05in} 30 &2\hspace{.05in} 2\hspace{.05in} 2 &Dec. 8,9 2002 \\
KMHK1042   &5 31 00 &-74 40 18 &15\hspace{.05in} 30\hspace{.05in} 30 &2\hspace{.05in} 2\hspace{.05in} 2 &Dec. 10 2002 \\
KMHK1278   &5 43 28 &-63 24 47 &15\hspace{.05in} 30\hspace{.05in} 30 &2\hspace{.05in} 2\hspace{.05in} 2 &Dec. 29 1998 \\
KMHK1381   &5 48 21 &-63 35 50 &15\hspace{.05in} 30\hspace{.05in} 30 &2\hspace{.05in} 2\hspace{.05in} 2 &Jan. 1 1999 \\
KMHK1339   &5 45 06 &-70 14 30 &15\hspace{.05in} 30\hspace{.05in} 30 &2\hspace{.05in} 2\hspace{.05in} 2 &Dec. 9 2002 \\
KMHK1640   &6 04 48 &-75 06 09 &15\hspace{.05in} 30\hspace{.05in} 30 &2\hspace{.05in} 2\hspace{.05in} 2 &Dec. 10 2002 \\
\hline
L11 (K7)   &0 27 45  & -72 46 53    & 10 \hspace{.05in} 15\hspace{.05in} 32.5 & 3\hspace{.05in} 3\hspace{.05in} 3 &Aug. 22-23 2001 \\
L17  (K13) &0 35 42  & -73 35  51   & 10 \hspace{.05in} 15\hspace{.05in} 32.5 & 3\hspace{.05in} 3\hspace{.05in} 3 &Aug. 22-23 2001\\
L113       &1 49 29  & -73 43  42   & 10 \hspace{.05in} 15\hspace{.05in} 32.5 & 3\hspace{.05in} 3\hspace{.05in} 3 &Aug. 22-23 2001 \\
NGC376     &1 03 50  & -72 49  34   & 10 \hspace{.05in} 15\hspace{.05in} 32.5 & 3\hspace{.05in} 3\hspace{.05in} 3 &Aug. 22-23 2001 \\
NGC419 (L85)&1 08 29 & -72 53  12   & 10 \hspace{.05in} 15\hspace{.05in} 32.5 & 3\hspace{.05in} 3\hspace{.05in} 3 &Aug. 22-23 2001  \\
NGC330       &0 56 19  &-72 27 50  & 10 \hspace{.05in} 15\hspace{.05in} 32.5 & 3\hspace{.05in} 3\hspace{.05in} 3 &Aug. 22-23 2001   \\
L80          &1 07 28  &-72 46 10  & 10 \hspace{.05in} 15\hspace{.05in} 32.5 & 3\hspace{.05in} 3\hspace{.05in} 3 &Aug. 22-23 2001  \\
NGC361       &1 02 11  &-71 36 21  & 10 \hspace{.05in} 15\hspace{.05in} 32.5 & 3\hspace{.05in} 3\hspace{.05in} 3 &Aug. 22-23 2001  \\
\hline
\end{tabular}
\label{Obsrerv}
\end{table*}

\begin{figure}
\centering
\includegraphics[width=6cm]{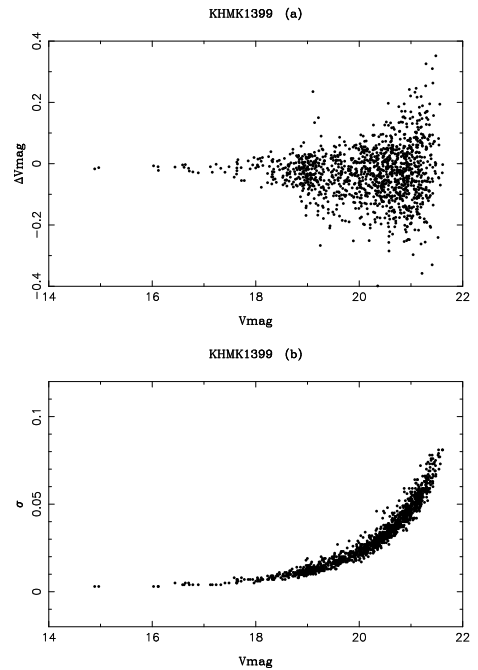}
\caption{$\bf a.$ The adopted difference in DAOPHOT mag within the two frames for the cluster KMHK1399 for the $V$ colour. $\bf b.$ The adopted standard error for the cluster KMHK1399 in the $V$ mag.} 
\label{DAO_errdiff_mag}
\end{figure}

In the LMC, two frames were available in each colour, and used to obtain the average magnitudes for the CMDs and 
$m_{1}$ (vs) $b-y$ diagrams. The adopted difference in DAOPHOT mag within  the two frames are shown in 
Fig.~\ref{DAO_errdiff_mag}a for the cluster KMHK1399 in the $V$ mag. A typical diagram of the standard 
error derived for the filter $V$ of cluster KMHK1399 is shown in Fig.~\ref{DAO_errdiff_mag}b.
For most of the SMC clusters three frames are used to derive the standard error for each filter.
However in the cases of L80, NGC330 and NGC361 only one frame was available for each filter.

An appropriate set of standard stars was obtained each night in order to achieve a reliable calibration. Transformations from the 
instrumental system to the standard system was obtained using the following equations (Richter et al. \cite{Richter}).

\hspace{-0.65cm}
y$_{inst}$ = V$_{st}$ + A$_{y}$ + B$_{y}\cdot$ X$_{y}$ + C$_{y}\cdot$ (b - y)$_{st}$ \\
b$_{inst}$ = b$_{st}$ + A$_{b}$ + B$_{b}\cdot$ X$_{b}$ + C$_{b}\cdot$ (b - y)$_{st}$ \\
v$_{inst}$ = v$_{st}$ + A$_{v}$ + B$_{v}\cdot$ X$_{v}$ + C$_{v}\cdot$ (v - b)$_{st}$ \\

\begin{table*}[h!]
\caption{Transformation coefficients}
\tiny
\begin{tabular}{lccccccccccccccc} 
\hline
\hline
Date        & A$_{y}$ & err    & B$_{y}$ & C$_{y}$ & err &  A$_{b}$ &  err &  B$_{b}$ & C$_{b}$&  err  &  A$_{v}$ &  err  & B$_{v}$ &  C$_{v}$ & err\\
28 Dec 1997 &  3.109  & 0.009  & 0.137  & -0.040  & 0.022  & 3.198  & 0.011  & 0.192  & -0.023  & 0.026 &  3.285  & 0.012 &  0.312 & -0.025  & 0.018\\
29 Dec 1997 &  3.136  & 0.002  & 0.111  & -0.047  & 0.008  & 3.224  & 0.003  & 0.165  & -0.027  & 0.009 &  3.343  & 0.005 &  0.263 & -0.026  & 0.010\\
01 Jan 1998 &  3.113  & 0.003  & 0.140  & -0.039  & 0.011  & 3.193  & 0.003  & 0.202  & -0.016  & 0.011 &  3.291  & 0.004 &  0.309 & -0.018  & 0.009\\
03 Jan 1998 &  3.153  & 0.002  & 0.120  & -0.034  & 0.007  & 3.228  & 0.004  & 0.190  & -0.005  & 0.010 &  3.319  & 0.005 &  0.300 & -0.024  & 0.009\\
14 Dec 1998 &  3.178  & 0.007  & 0.108  & -0.033  & 0.006  & 2.484  & 0.008  & 0.165  & -0.056  & 0.007 &  2.671  & 0.013 &  0.288 &  0.009  & 0.008\\
16 Dec 1998 &  3.165  & 0.002  & 0.129  & -0.019  & 0.006  & 2.480  & 0.002  & 0.180  & -0.045  & 0.006 &  2.682  & 0.005 &  0.291 &  0.022  & 0.007\\
08 Dec 2002 &  2.456  & 0.002  & 0.131  & -0.044  & 0.006  & 2.382  & 0.003  & 0.192  & -0.035  & 0.007 &  2.379  & 0.005 &  0.306 & -0.003  & 0.008\\
09 Dec 2002 &  2.466  & 0.005  & 0.131  & -0.041  & 0.012  & 2.398  & 0.004  & 0.192  & -0.038  & 0.009 &  2.394  & 0.009 &  0.306 & -0.010  & 0.011\\
10 Dec 2002 &  2.468  & 0.011  & 0.131  & -0.002  & 0.024  & 2.401  & 0.004  & 0.192  & -0.051  & 0.010 &  2.336  & 0.007 &  0.306 & -0.007  & 0.009\\
22 Aug 2001 &  0.832  & 0.005  & 0.145  & -0.019  & 0.015  & 0.658  & 0.005  & 0.209  &  0.020  & 0.017 &  0.374  & 0.017 &  0.330 &  0.032  & 0.026\\ 
23 Aug 2001 &  0.856  & 0.004  & 0.140  & -0.030  & 0.011  & 0.665  & 0.006  & 0.230  & -0.015  & 0.016 &  0.408  & 0.020 &  0.320 &  0.042  & 0.021\\
\hline
\hline
\end{tabular}
\label{tra_coeff}
\end{table*}

The B$_{y}$, B$_{b}$ and B$_{v}$ parameters are the atmospheric extinction coefficients for the  y, b, and v filters.
The X$_{y}$, X$_{b}$ are X$_{v}$ parameters are the AirMass at the three filters and they are known from the observations.
The B$_{y}$, B$_{b}$ and B$_{v}$ are also known, thus they are kept constant in the equations and they dont have any errors.
Finally using least square fittings we estimate the rest 6 parameters:
A$_{y}$, A$_{b}$, $A_{v}$, C$_{y}$, C$_{b}$, and C$_{v}$.
The values of the transformation coefficients are shown in Table~\ref{tra_coeff}.


The adopted criteria for the photometry to produce the CMDs are: 
a) During cross identification of stars on all available frames in all filters, only those stars with coordinates 
matching to better than 1 pixel (0.4 arcsec/pixel for the LMC and 0.32 arcsec/pixel for the SMC) were accepted. 
b) Photometric error for $y, b, v$ is found as the weighted average of the values found in the  corresponding frames.
We used these errors to determine the final errors in $b-y$ and $m_{1}$.
c) The stars adopted for the production of the CMDs are only those with error $0.1mag$ in $V$, $b-y$ and $m_{1}$. 
d) Using DAOPHOT we adopted as goodness of the PSF fit x$^{2}<1.9$ and image sharpness, $s$, $|s|<1$.

\section{Discussion}

\begin{figure} 
\centering
\includegraphics[width=8cm,trim=1.5cm 4.5cm 2cm 4cm]{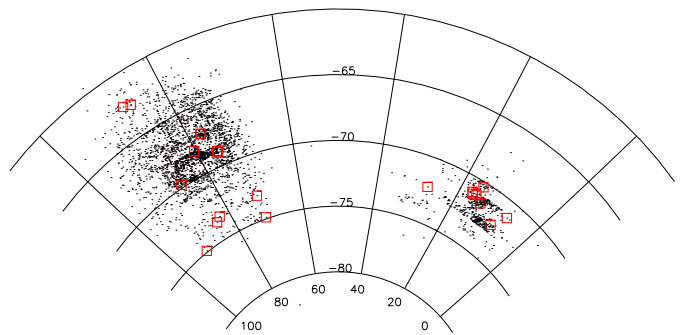}
\caption{The spatial distribution of the MCs star clusters under investigation.}
\label{f02}
\end{figure}

The present sample of clusters includes seven clusters (KMK1, KMK3, KMK8, HS153, KMK32, KMK49, KMK50) 
located in the central region of LMC and eight clusters (KMHK81, KMHK1042, KMHK1278, KMHK1381, KMHK1399, 
KMHK1640, SL36 and SL620) located in the outer region of the LMC. From the outermost clusters 3 are 
located in the north and 5 in the south with an average distance R $\le$ 6-7 Kpc from the centre. 
Generally the young  LMC clusters (a few $\times10^{8}$yr) are only located in the central region, 
whereas all other older ones are found all over the LMC (Kontizas et al. \cite{Kontizas90}). 
The SMC clusters are mostly chosen to be at the outskirts of the galaxy to avoid crowded regions. 
The spatial distribution of the MCs selected star clusters is shown in Fig.\ref{f02}. 
They are overplotted on the catalogue of the MCs star clusters of Bica et al. \cite{Bica08}.

Str\"{o}mgren photometry is known to provide an excellent metallicity indicator for late
type stars (Richtler, \cite{Richtler88}, \cite{Richtler89},  Grebel \& Richler, \cite{Grebel_R_92}). 
It is particularly efficient when it is performed with a CCD in dense star
fields like the MCs. The validity domain in case of giants
and red supergiants is $0.4<b-y<1.1$ (Grebel \& Richler, \cite{Grebel_R_92}, Hilker et al.,\cite{Hilker95}).
On the Str\"{o}mgren V, (b-y) Color Magnitude Diagram (CMD) we fit the isochrone that best describes  
the stellar content of the cluster. From the isochrone we derive age and an estimation of metallicity. Then we search
for red supergiants. We compare with red supergiants of the field. The
ideal case is when we can find red supergiants of the cluster that do not
appear in the field. Then we trace the position of these stars on the $m_{1}$, $(b-y)$
diagram and compare with the model lines that indicate equal
metallicity. Thus we provide the estimation of the Str\"{o}mgren metallicity.

More details of the procedure are given in the following subsections.

\subsection{Ages of the clusters}

For each cluster we produced a $V, (b-y)$ CMD. In order to trace the cluster 
stars among the contaminated nearby and/or projected field stars, we carried out the following: A central 
region around each cluster centre was chosen, as small as possible ($\sim$0.75arcmin radius), in order 
to include the largest proportion of cluster members and large enough, because sometimes crowding was too severe 
to have measurements of the very central stars. In the outer parts of the cluster we were able to find a region 
characterizing the nearby field stellar population. We selected such fields with equal area with that of the cluster 
and produced the corresponding CMDs. Comparison of the two diagrams (central cluster \& field respectively) may allow 
the determination of the cluster members. In case there is a significant difference of their CMDs, the determination 
of the cluster parameters such as age and metallicity is much more reliable than in cases, where the two diagrams have 
small differences. 

Then we fit the isochrone that best describes  
the stellar population of the cluster. The models used are those of Schaerer et al. 
\cite{Schaerer93a}; Schaerer et al. \cite{Schaerer93b}; Schaller et al. \cite{Schaller92} and Charbonnel et al. 
\cite{Charbonnel93} with an appropriate transformation for the Str\"{o}mgren magnitudes.
The isochrones provide the parameters of age in Gyr and metalicity Z.
The E(b-y) is determined by estimating how much one should  move the isochrones on the red (right direction) in order 
to match better with the stars.

In Fig.~\ref{f03} to Fig.~\ref{f17} and Fig.~\ref{f18} to Fig.~\ref{f25} the CMDs of the studied 
clusters and their adjoining fields are given for the LMC and SMC respectively.
In each figure the upper two diagrams show the $V, (b-y)$ CMDs for the cluster and the field respectively.
In the upper left diagram the isochrone that best describes the cluster population is overplotted. The derived values of 
the age, metallicity and the extinction for each cluster are listed in Table~\ref{ages}, columns 2, 4 and 5 respectively.
The metallicity values are transformed from Z to [Fe/H], using the transformation table by Durand et al., \cite{Durand84}.

\subsection{Metallicities}

\begin{figure} 
\centering
\includegraphics[width=7.5cm]{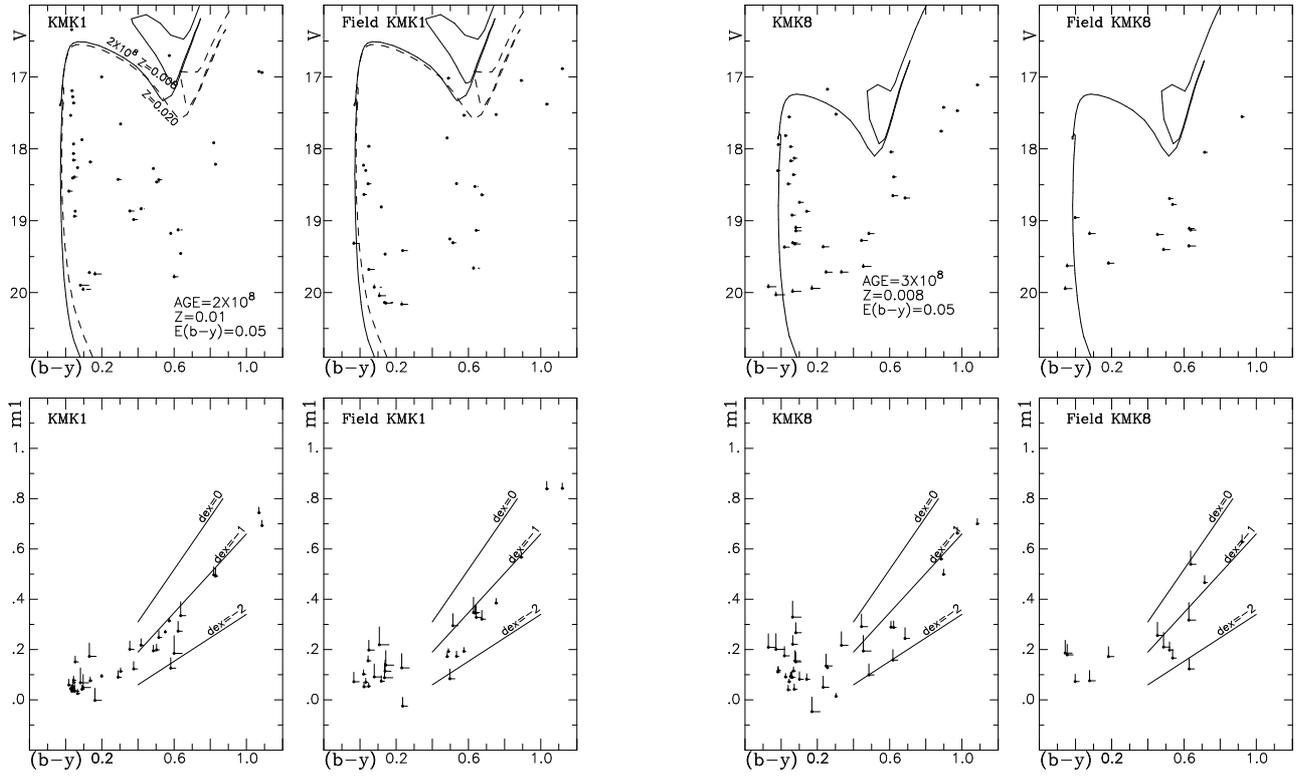}
\tiny
\caption{CMD for the cluster KMK1 and its, equal area, field for r=0.75arcmin.  Metallicity for the cluster KMK1 and its field.}
\label{f03}
\end{figure}

\begin{figure} 
\centering
\includegraphics[width=7.5cm]{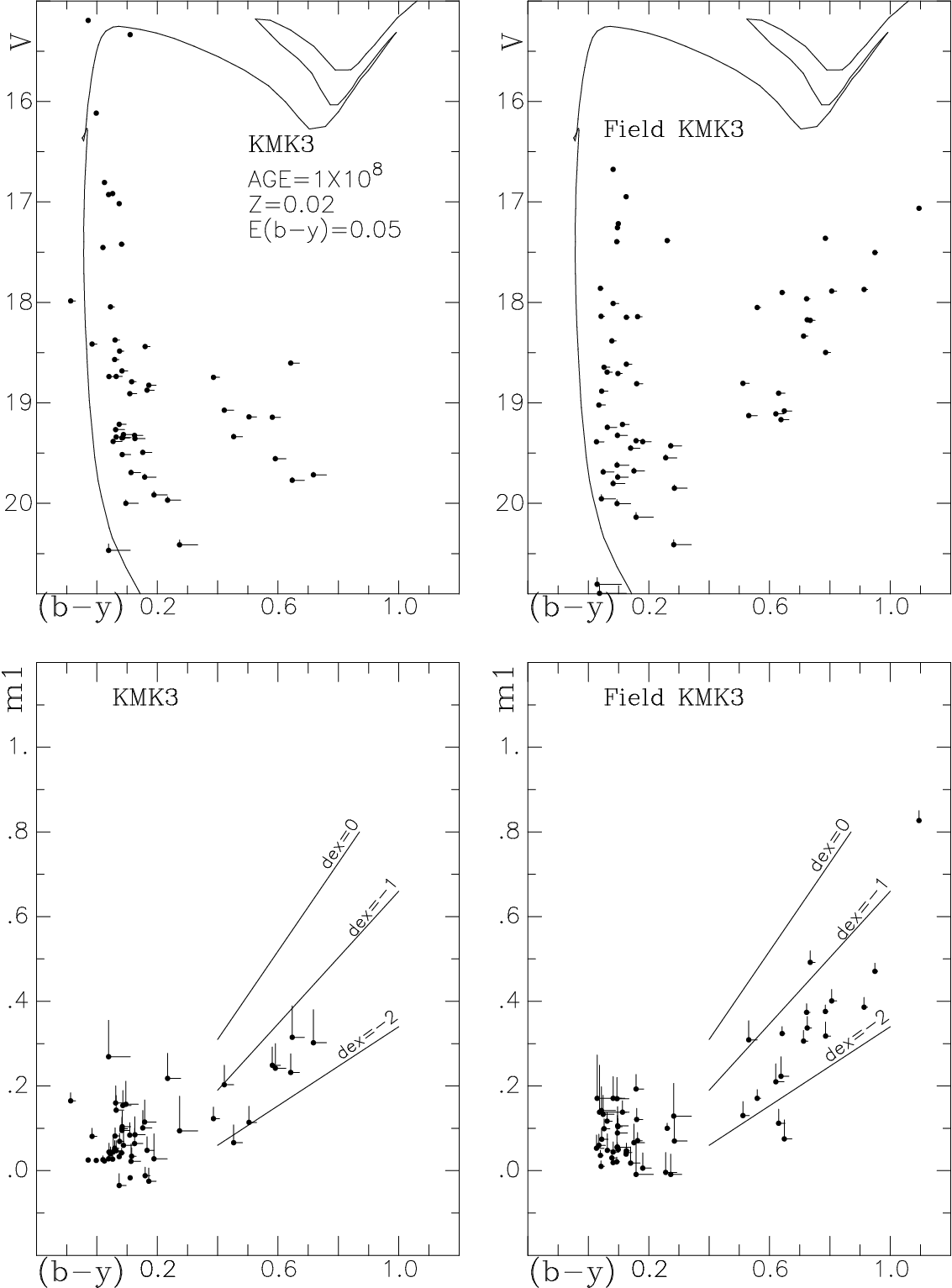}
\tiny
\caption{Same as Fig. ~\ref{f03}  for the cluster KMK3 and its, adjoining field.}
\label{f04}
\end{figure}

\begin{figure} 
\centering
\includegraphics[width=7.5cm]{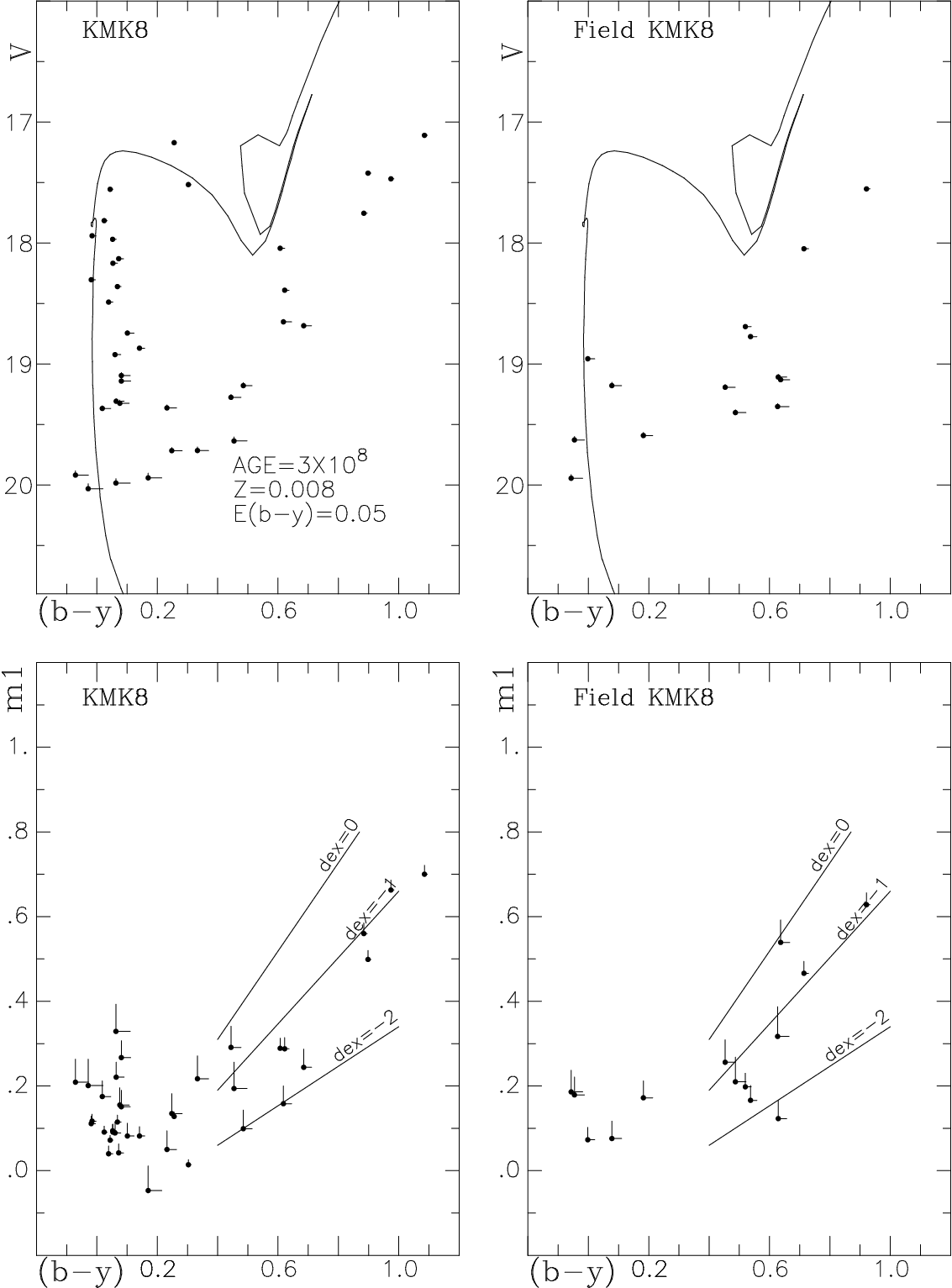}
\tiny
\caption{Same as Fig. ~\ref{f03}  for the cluster KMK8 and its, adjoining field.}
\label{f05}
\end{figure}

\begin{figure} 
\centering
\includegraphics[width=7.5cm]{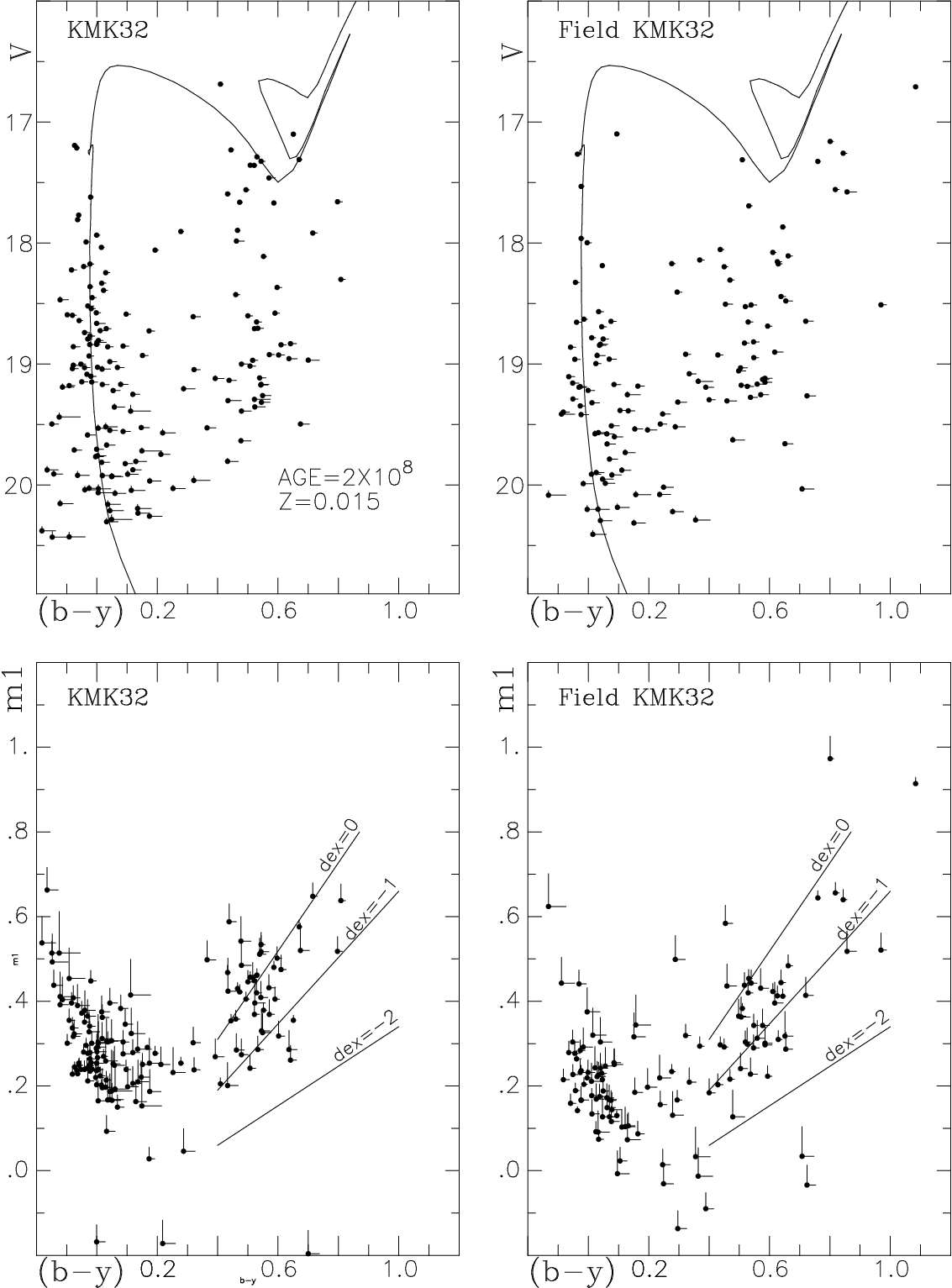}
\tiny
\caption{Same as Fig. ~\ref{f03}  for the cluster KMK32 and its, adjoining field.}
\label{f06}
\end{figure}

\begin{figure} 
\centering
\includegraphics[width=7.5cm]{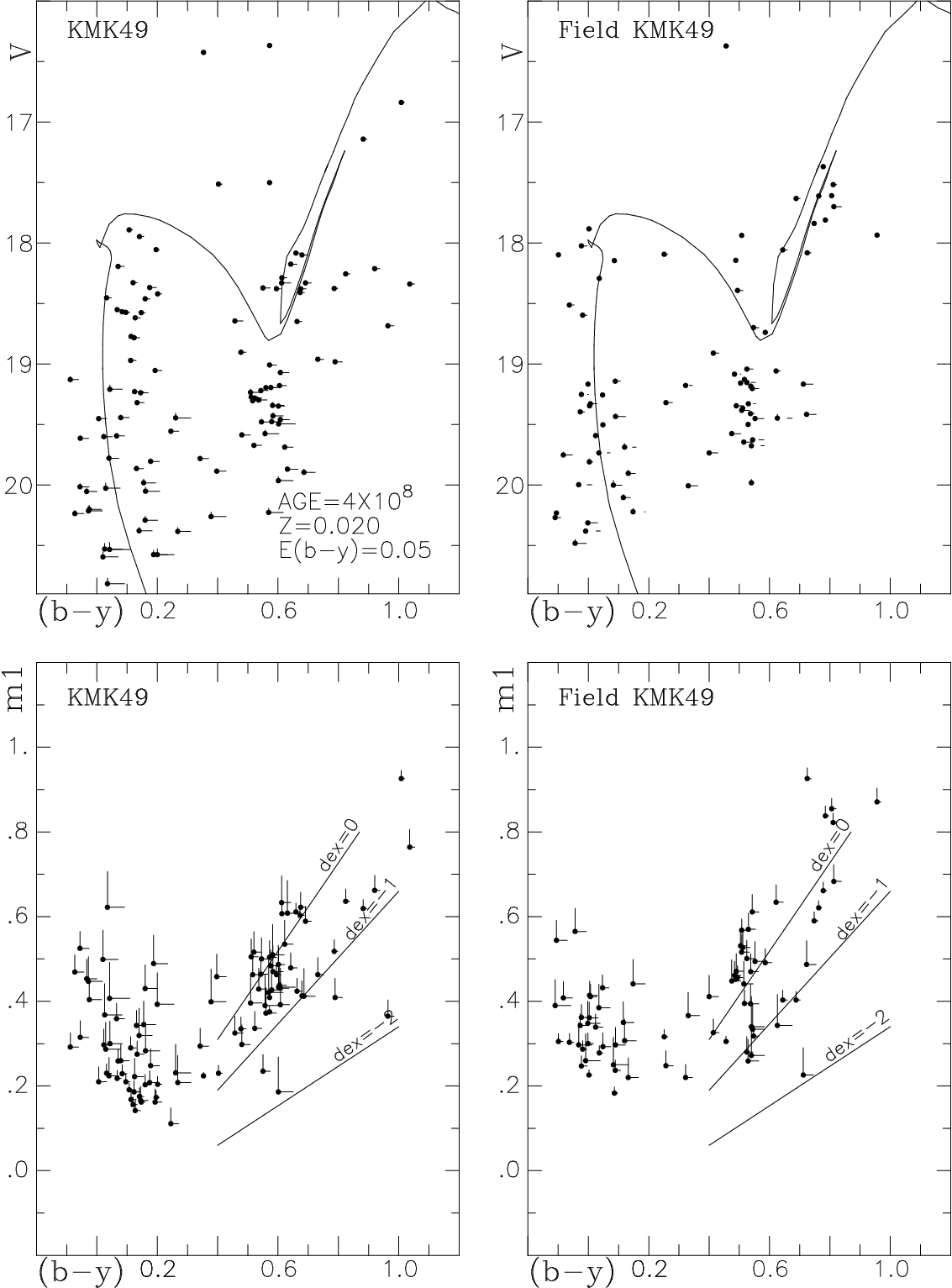}
\tiny
\caption{ Same as Fig. ~\ref{f03}  for the cluster KMK49 and its, adjoining field.}
\label{f07}
\end{figure}

\begin{figure} 
\centering
\includegraphics[width=7.5cm]{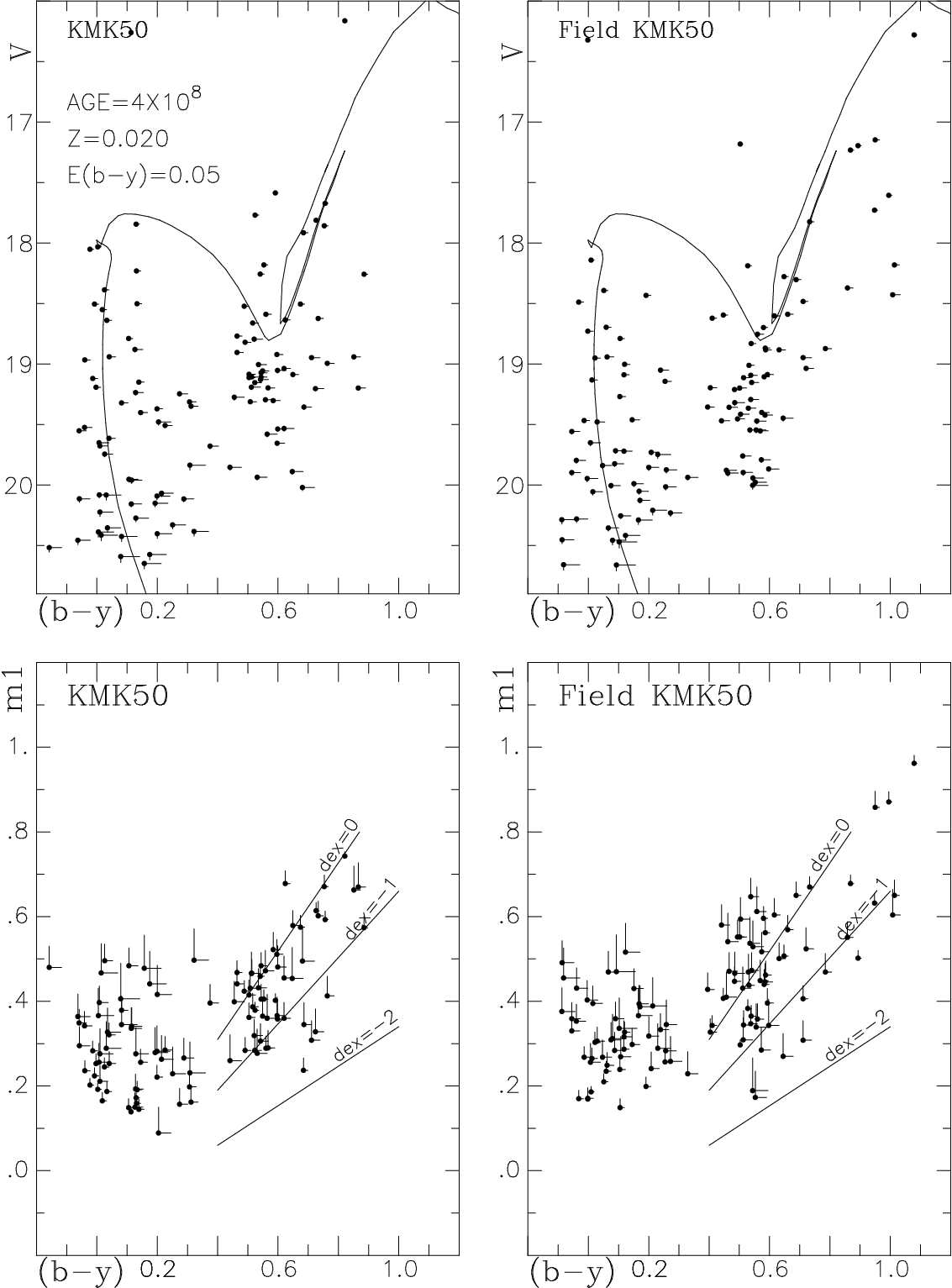}
\tiny
\caption{ Same as Fig. ~\ref{f03}  for the cluster KMK50 and its, adjoining field.}
\label{f08}
\end{figure}

\begin{figure} 
\centering
\includegraphics[width=7.5cm]{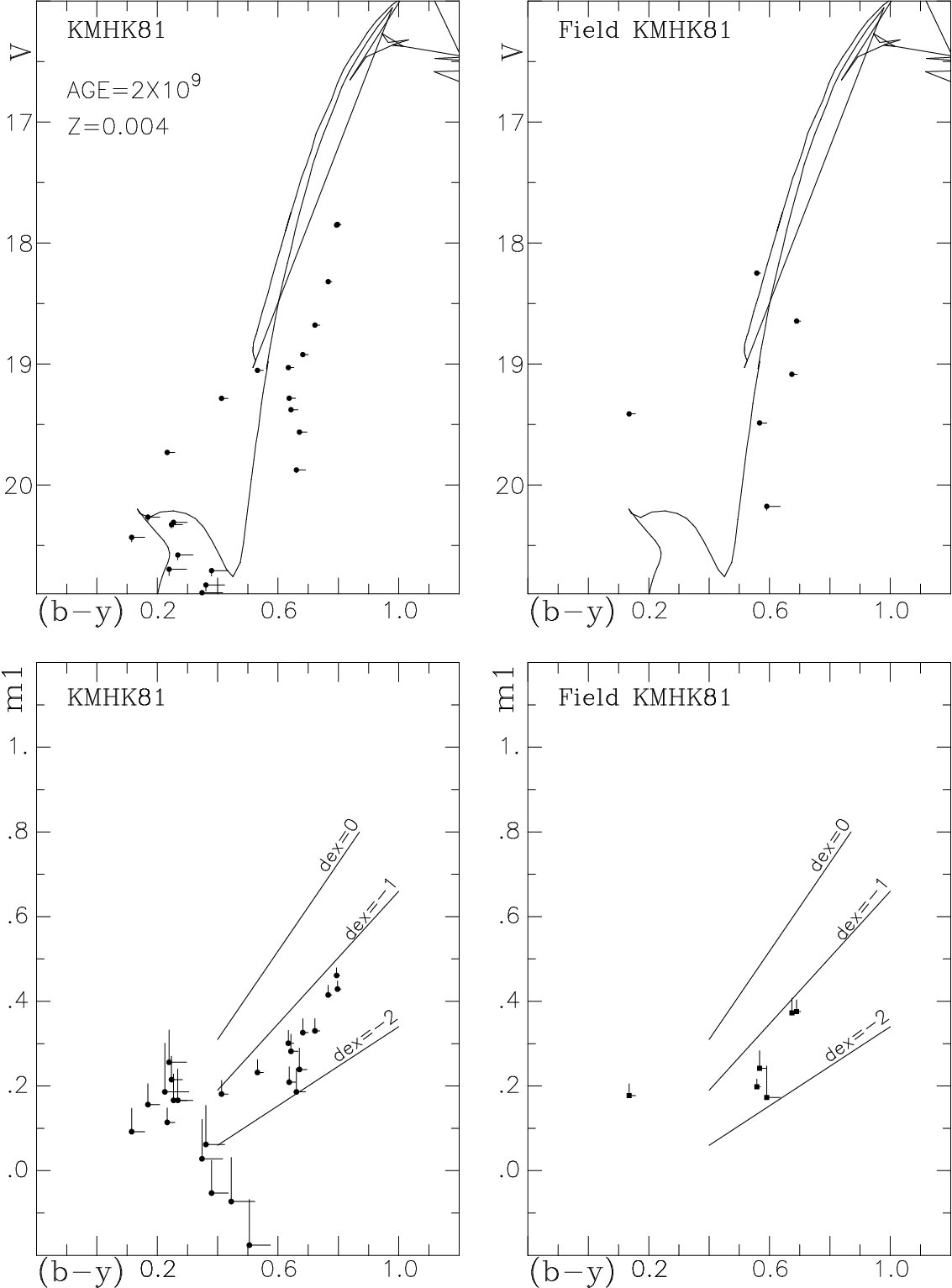}
\tiny
\caption{ Same as Fig. ~\ref{f03}  for the cluster KMK81 and its, adjoining field.}
\label{f09}
\end{figure}

\begin{figure} 
\centering
\includegraphics[width=7.5cm]{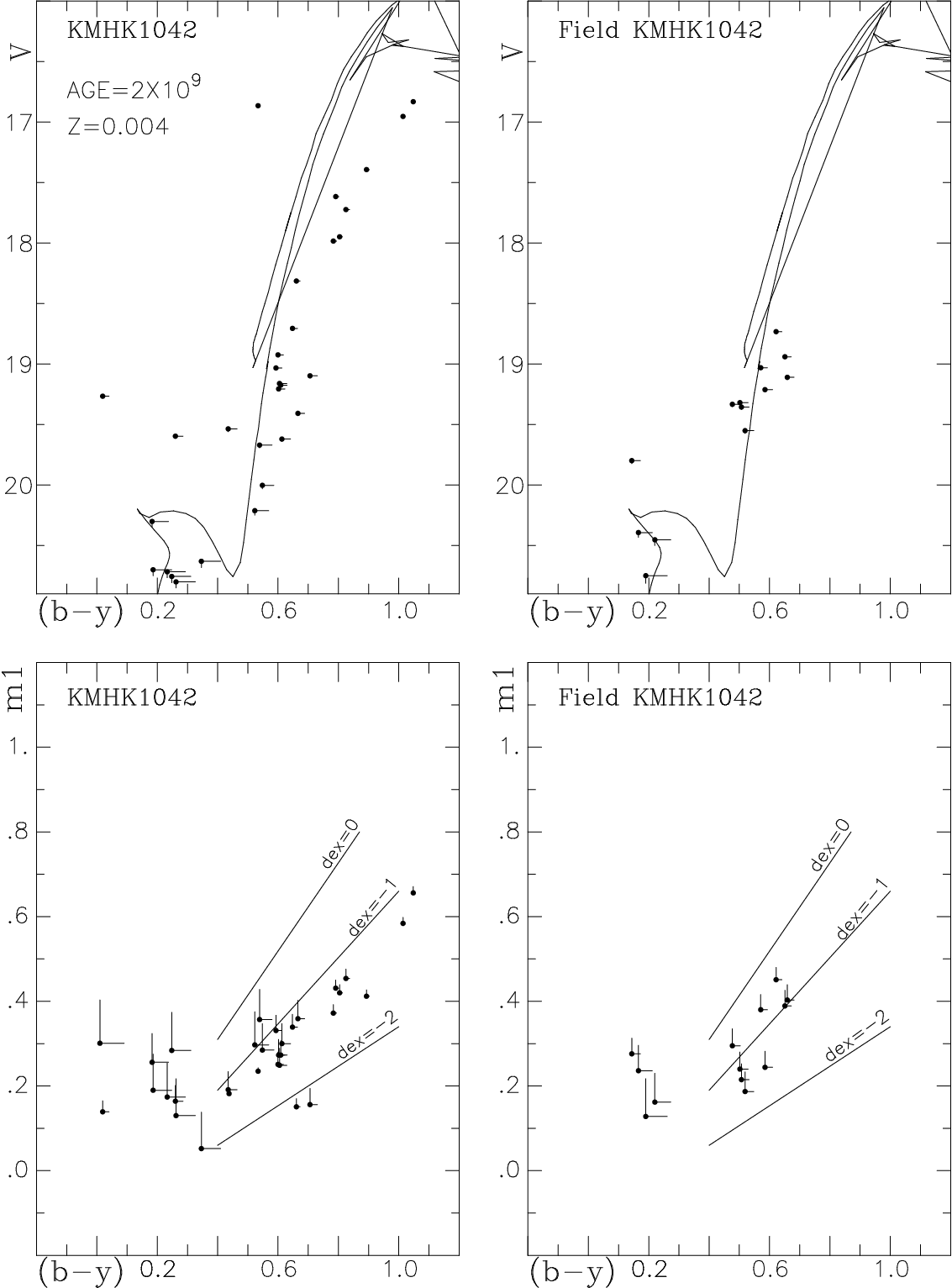}
\tiny
\caption{ Same as Fig. ~\ref{f03}  for the cluster KMHK1042 and its, adjoining field.}
\label{f10}
\end{figure}

\begin{figure} 
\centering
\includegraphics[width=7.5cm]{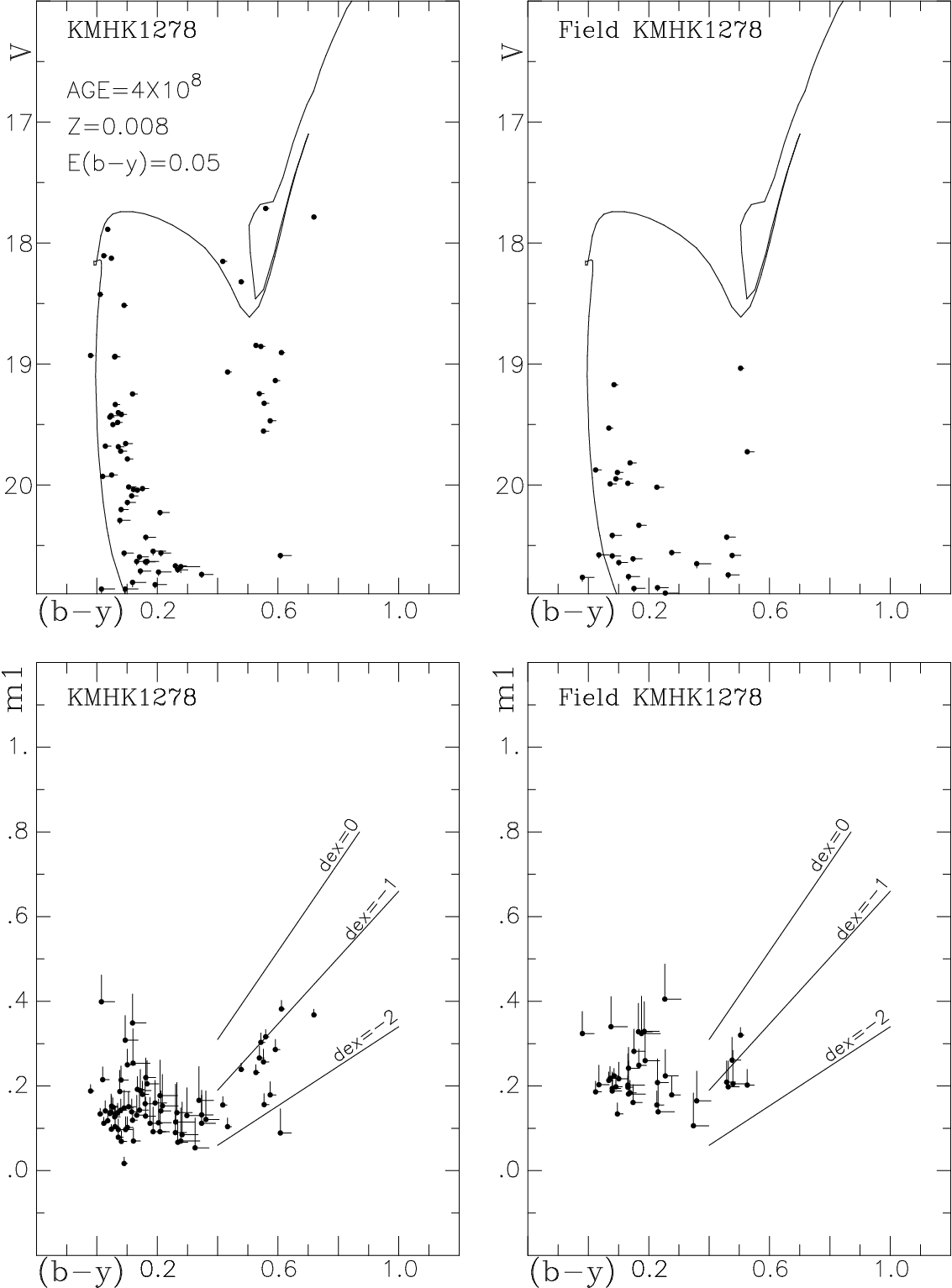}
\tiny
\caption{ Same as Fig. ~\ref{f03}  for the cluster KMHK1278 and its, adjoining field.}
\label{f11}
\end{figure}

\begin{figure} 
\centering
\includegraphics[width=7.5cm]{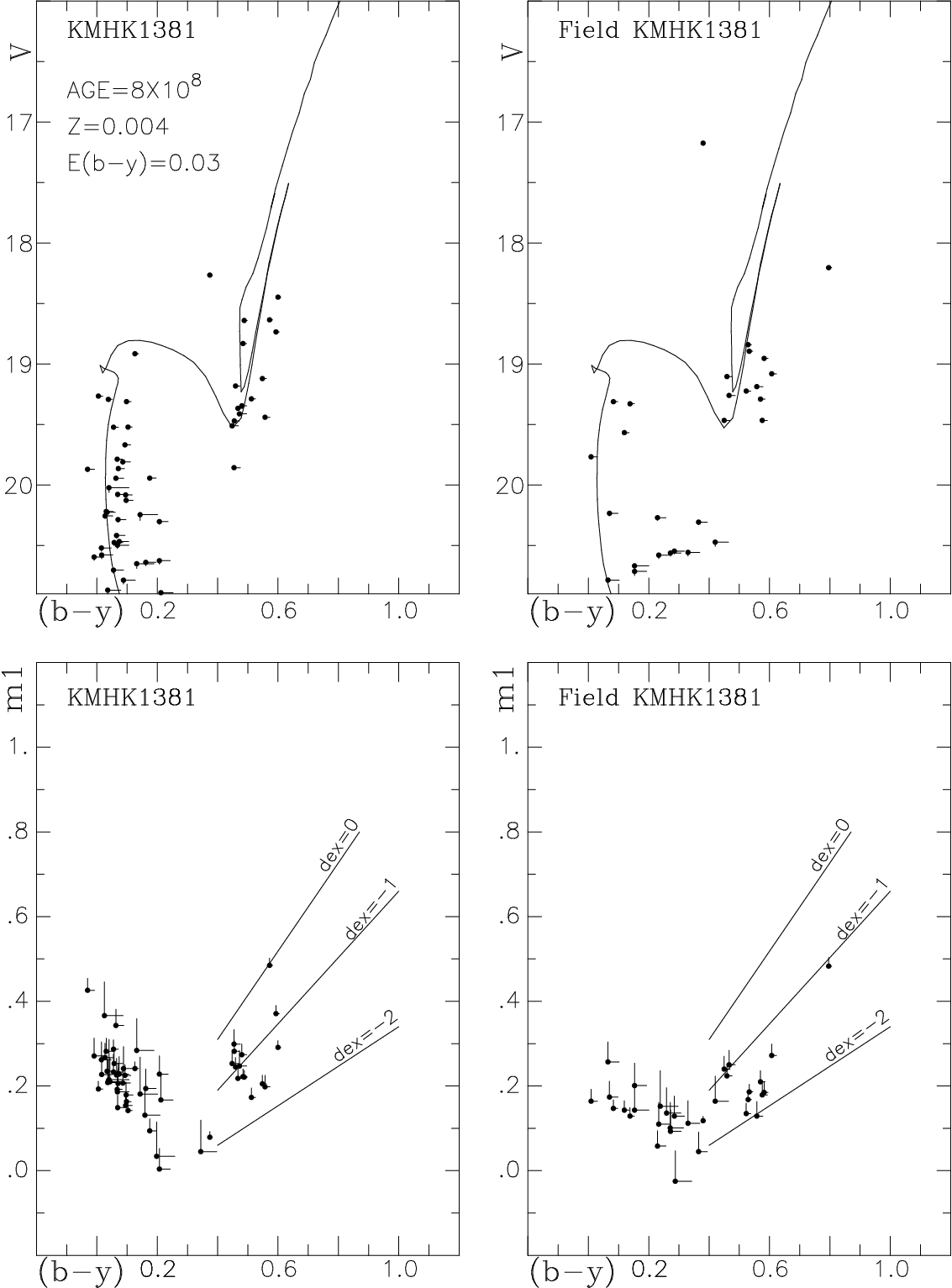}
\tiny
\caption{ Same as Fig. ~\ref{f03}  for the cluster KMHK1381 and its, adjoining field.}
\label{f12}
\end{figure}

\begin{figure} 
\centering
\includegraphics[width=7.5cm]{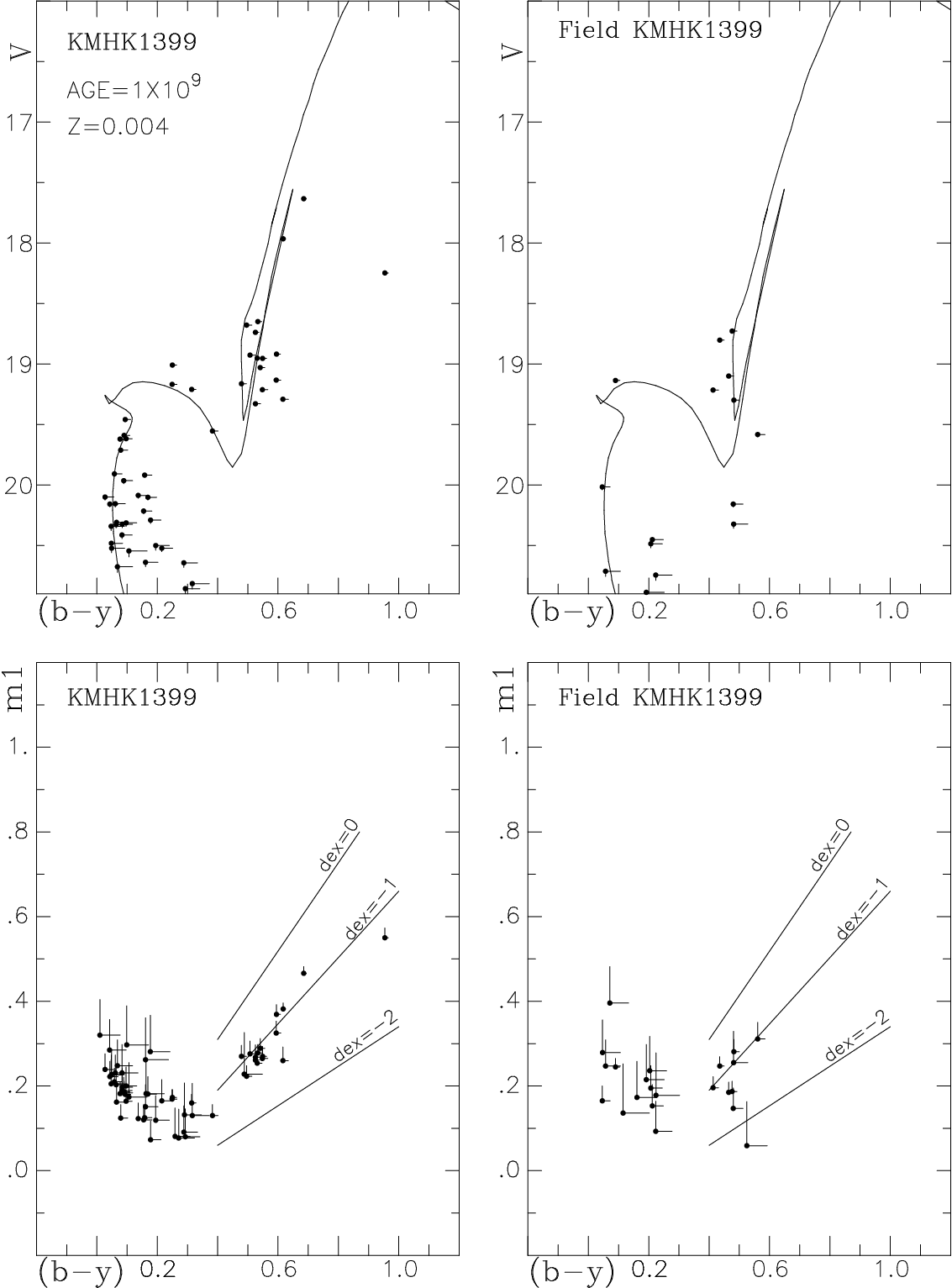}
\tiny
\caption{ Same as Fig. ~\ref{f03}  for the cluster KMHK1399 and its, adjoining field.}
\label{f13}
\end{figure}

\begin{figure} 
\centering
\includegraphics[width=7.5cm]{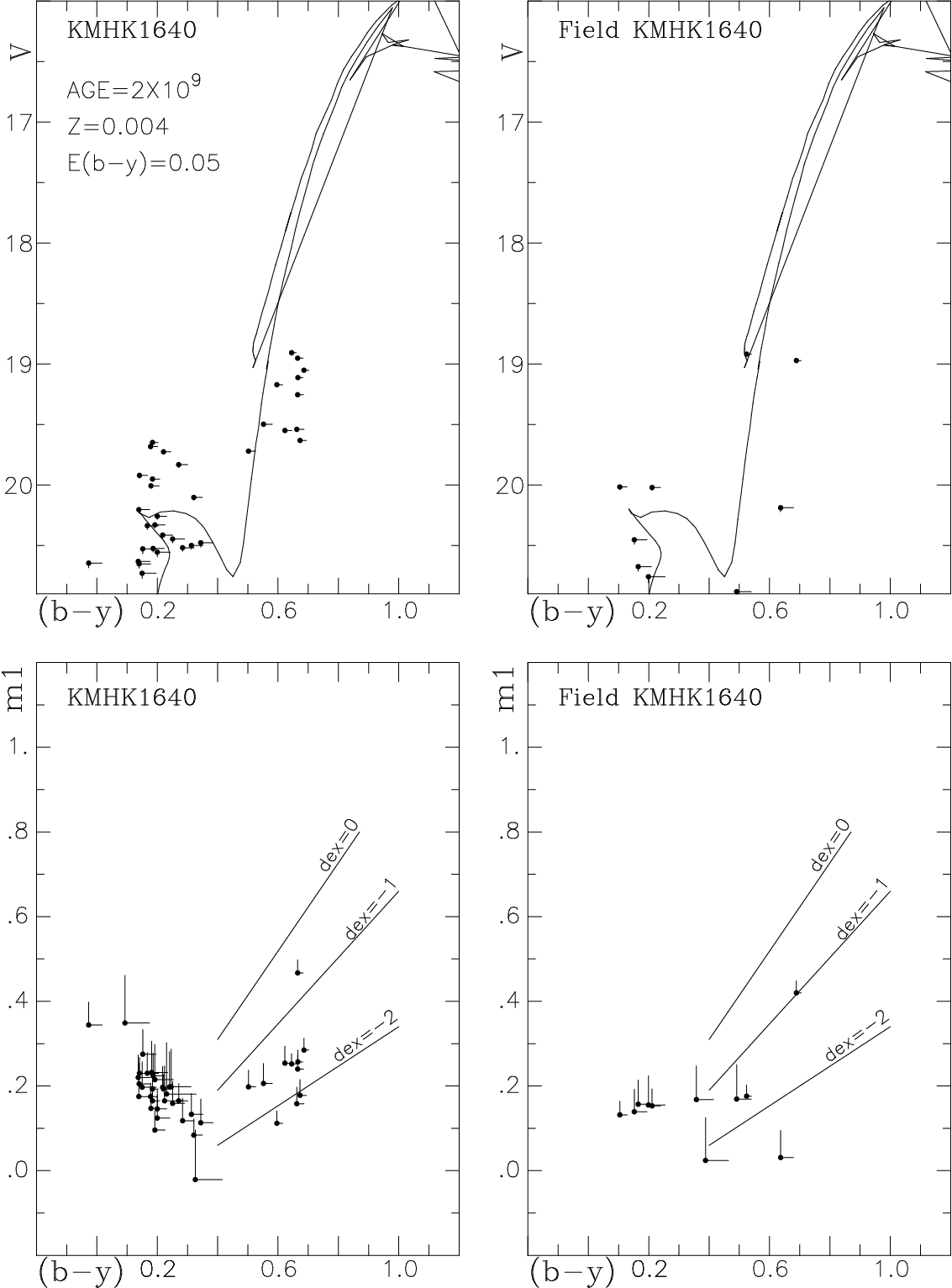}
\tiny
\caption{ Same as Fig. ~\ref{f03}  for the cluster KMHK1640 and its, adjoining field.}
\label{f14}
\end{figure}

\begin{figure} 
\centering
\includegraphics[width=7.5cm]{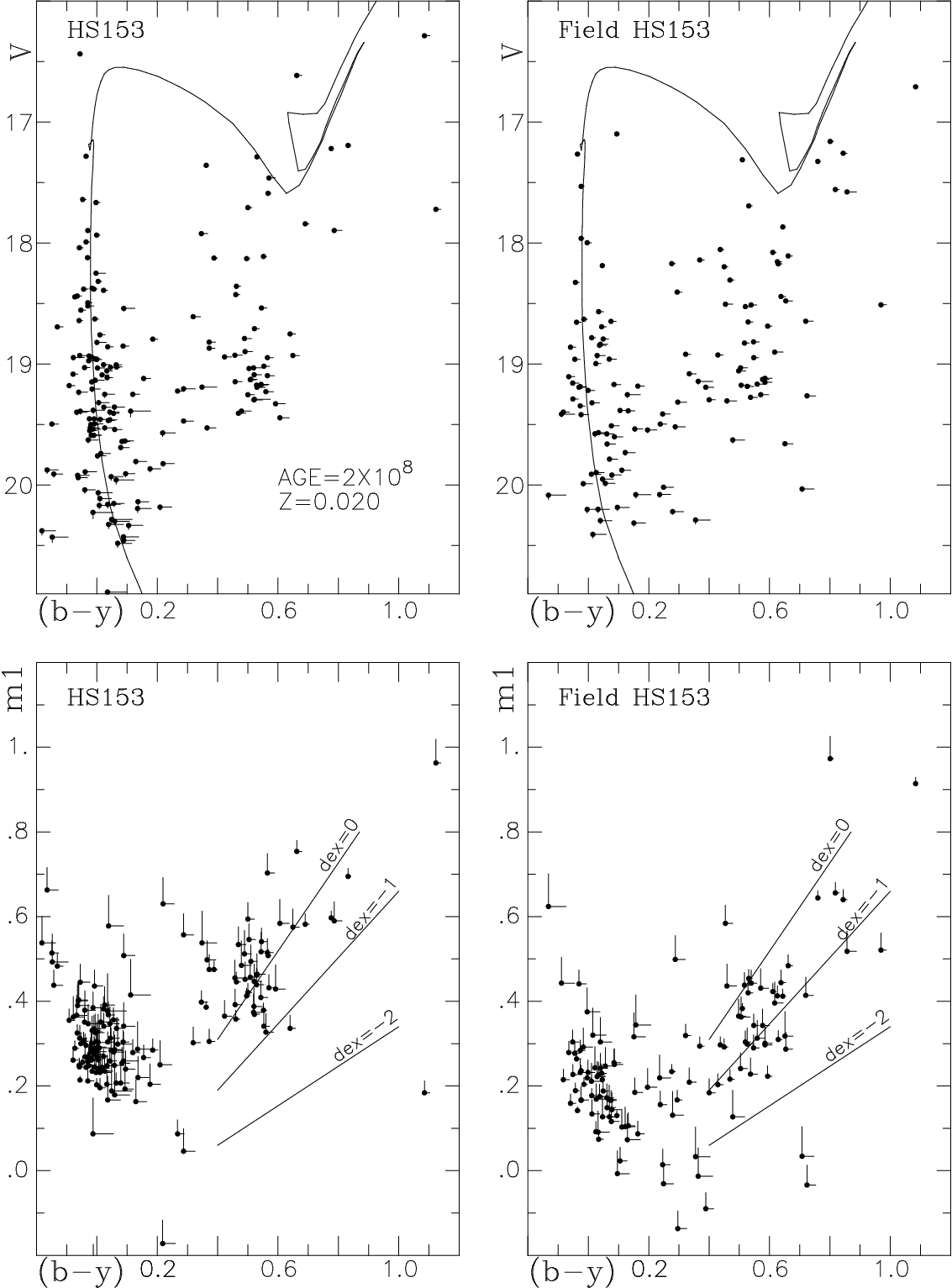}
\caption{ Same as Fig. ~\ref{f03}  for the cluster HS153 and its, adjoining field.}
\label{f15}
\end{figure}

\begin{figure} 
\centering
\includegraphics[width=7.5cm]{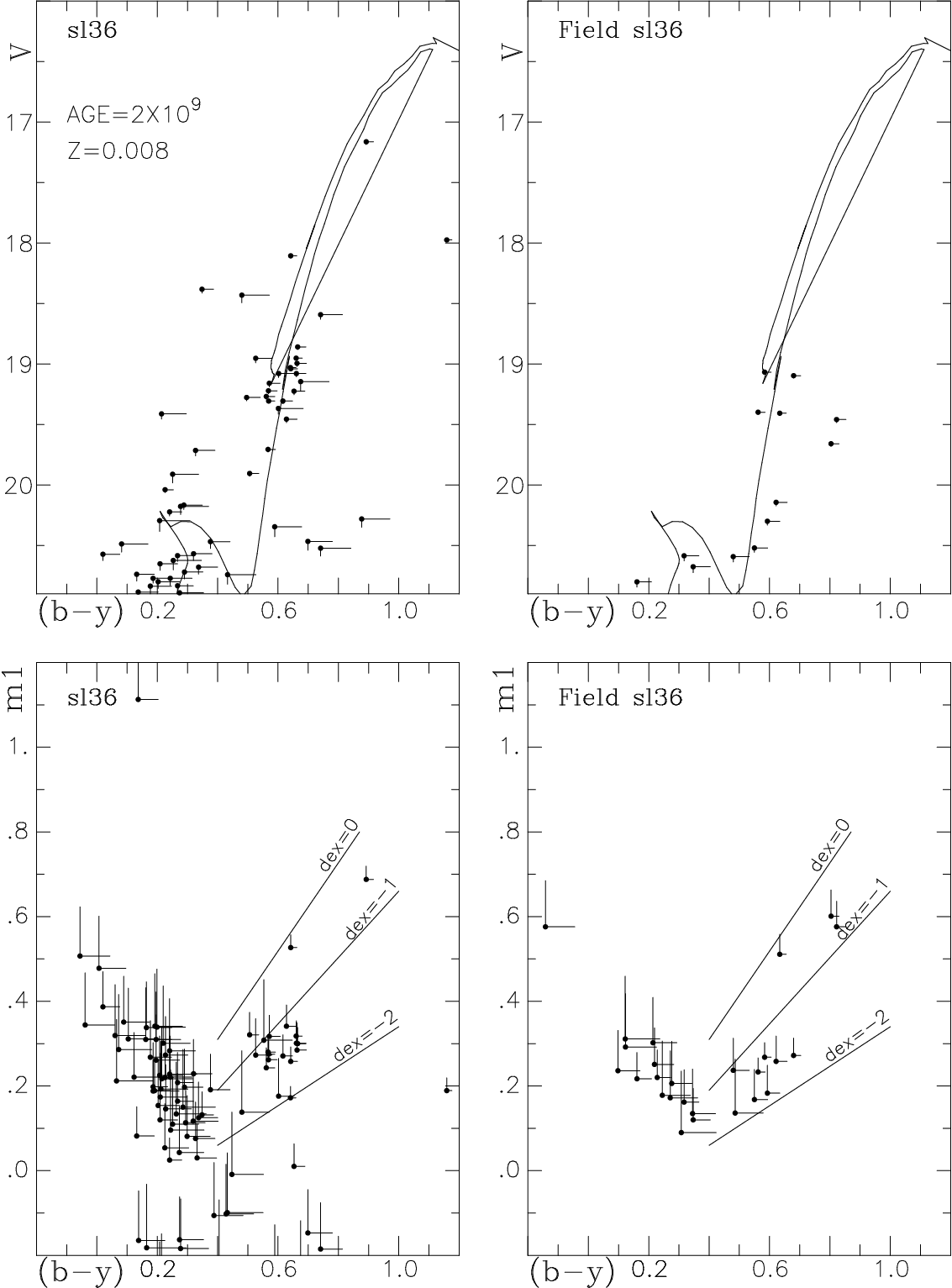}
\tiny
\caption{ Same as Fig. ~\ref{f03}  for the cluster SL36 and its, adjoining field.}
\label{f16}
\end{figure}

\begin{figure} 
\centering
\includegraphics[width=7.5cm]{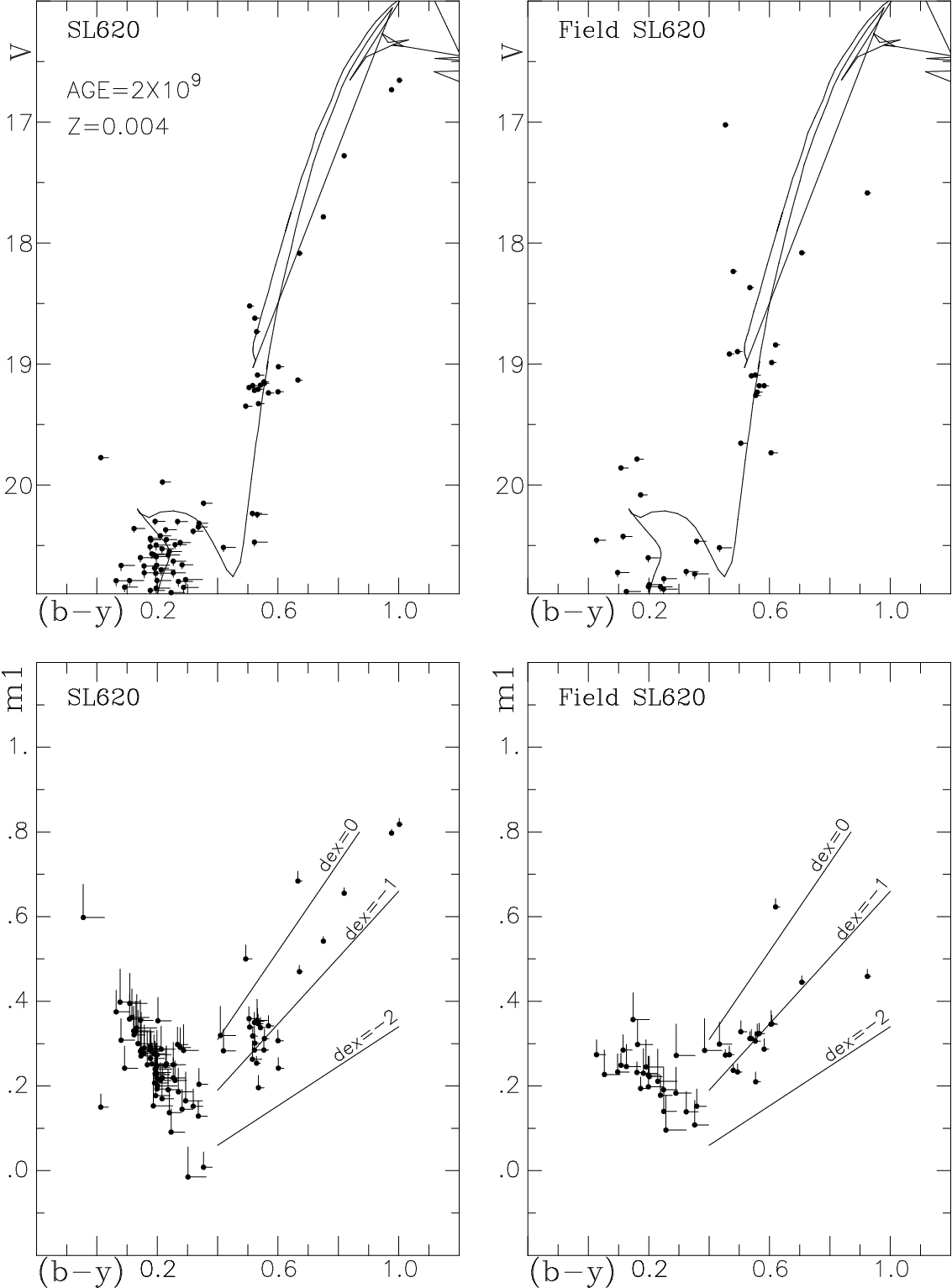}
\tiny
\caption{ Same as Fig. ~\ref{f03}  for the cluster SL620 and its, adjoining field.}
\label{f17}
\end{figure}

\begin{figure} 
\centering
\includegraphics[width=7.5cm]{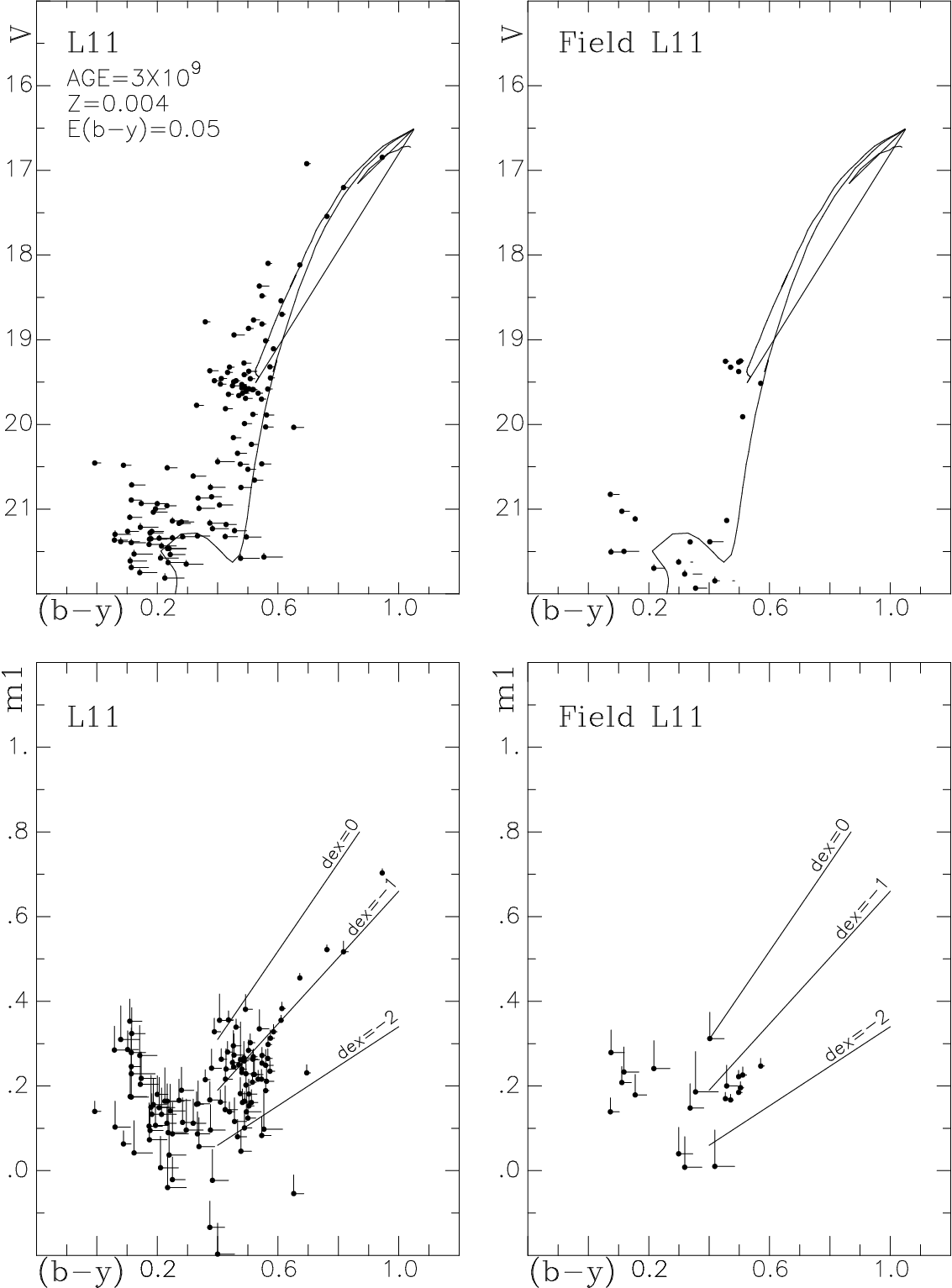}
\tiny
\caption{Same as Fig. ~\ref{f03}  for the cluster L11 and its, adjoining field.}
\label{f18}
\end{figure}

\begin{figure} 
\centering
\includegraphics[width=7.5cm]{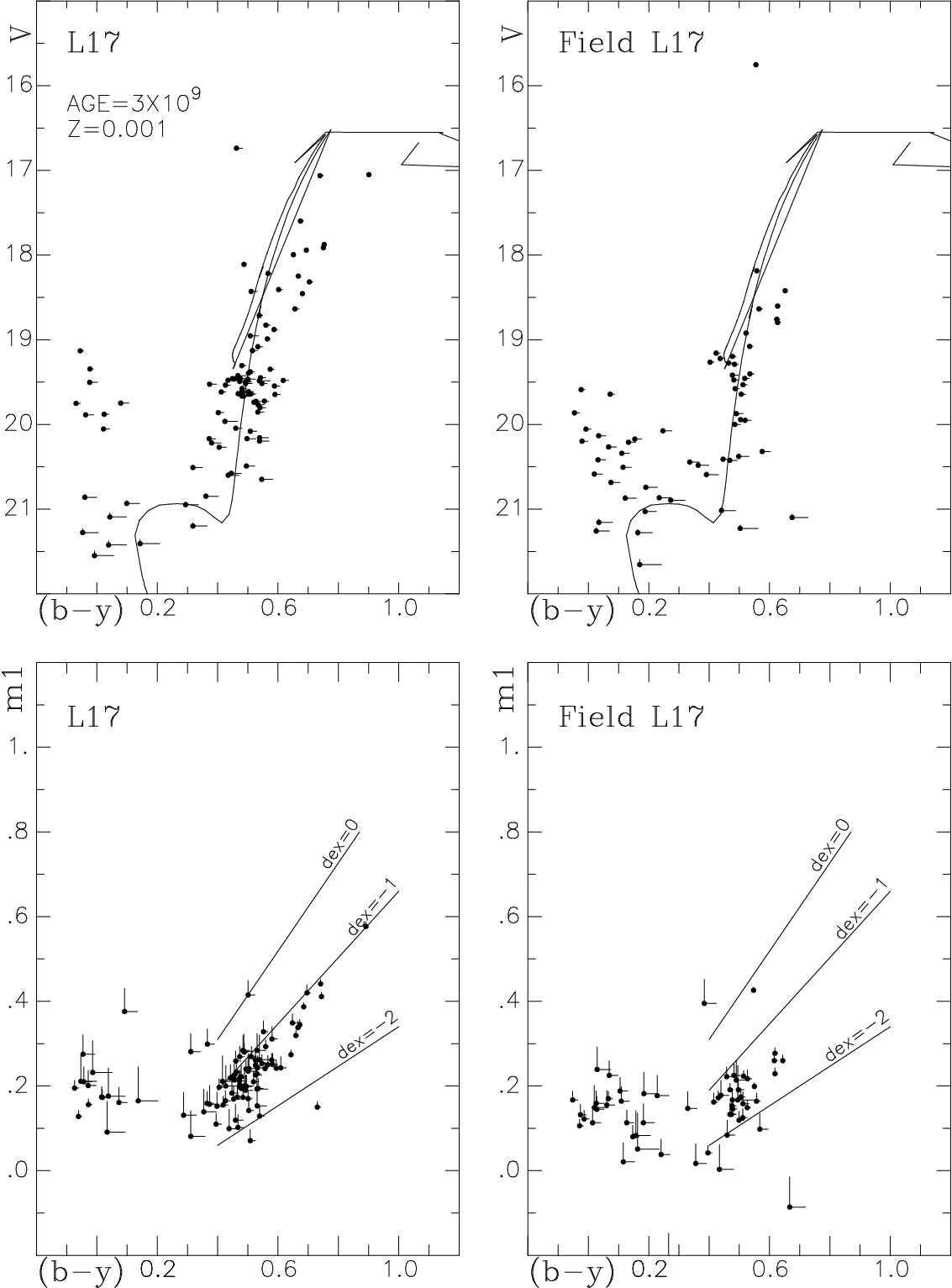}
\tiny
\caption{ Same as Fig. ~\ref{f03}  for the cluster L17 and its, adjoining field.}
\label{f19}
\end{figure}

\begin{figure} 
\centering
\includegraphics[width=7.5cm]{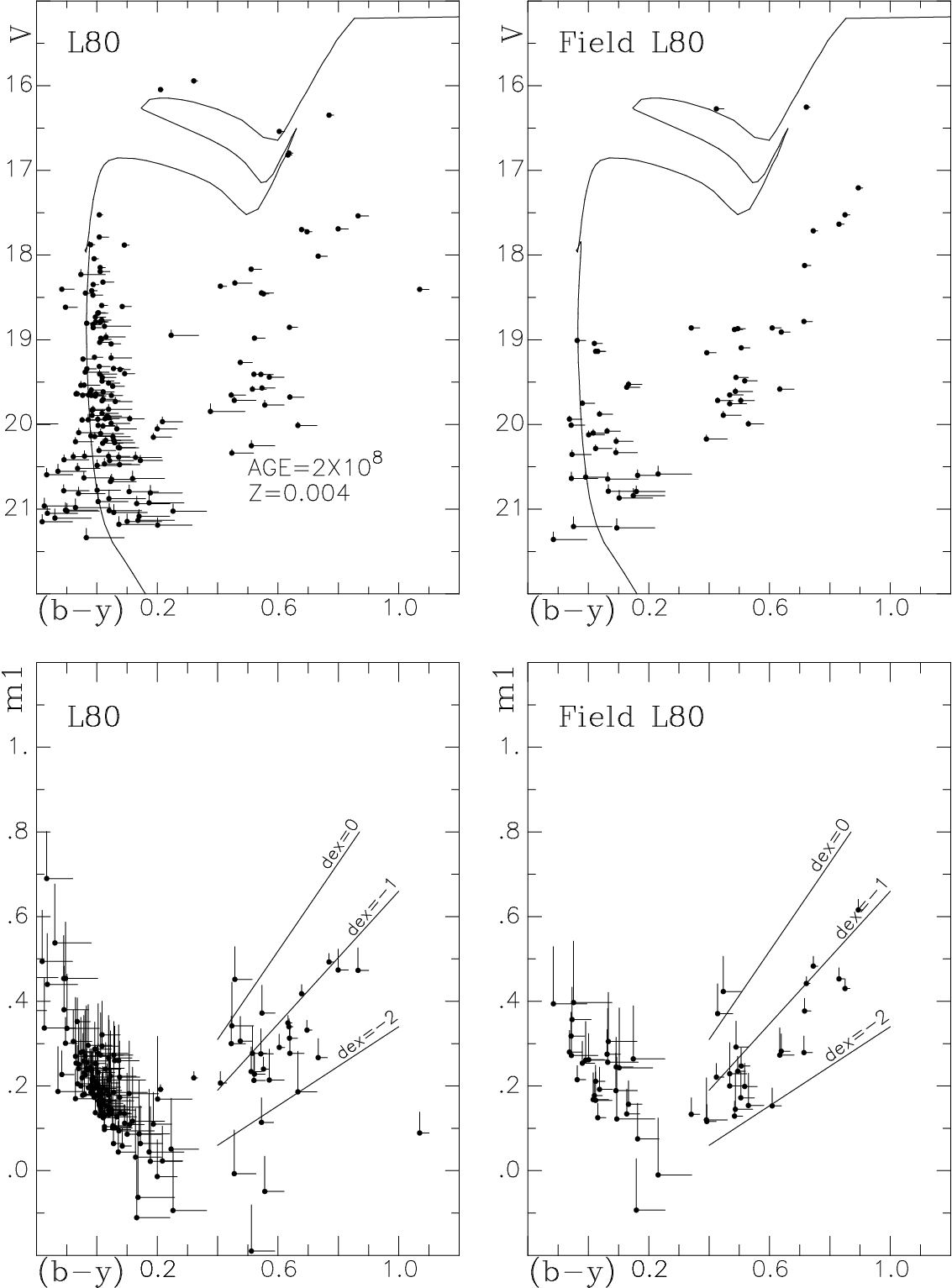}
\tiny
\caption{ Same as Fig. ~\ref{f03}  for the cluster L80 and its, adjoining field.}
\label{f20}
\end{figure}

\begin{figure} 
\centering
\includegraphics[width=7.5cm]{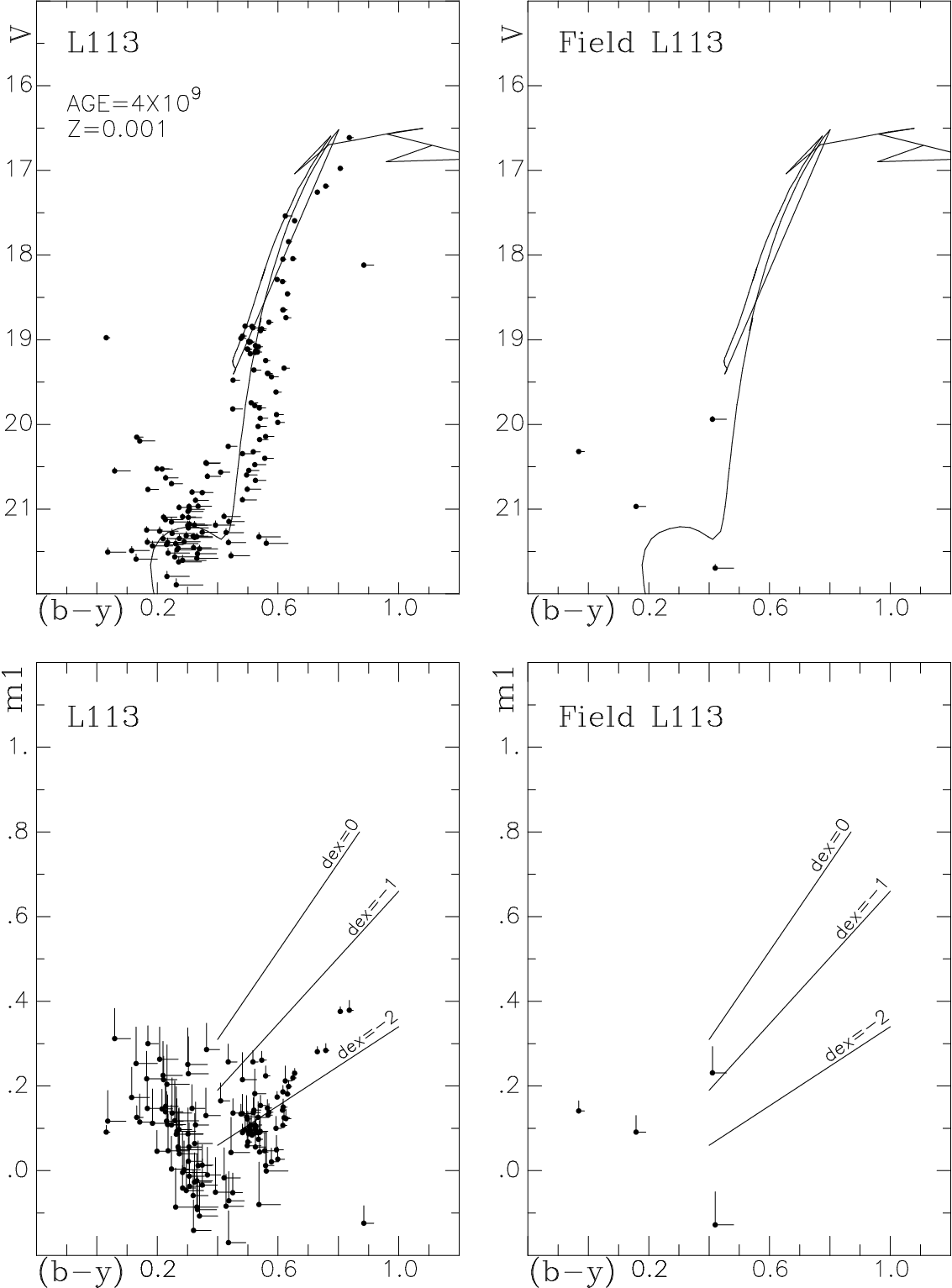}
\tiny
\caption{ Same as Fig. ~\ref{f03}  for the cluster L113 and its, adjoining field.}
\label{f21}
\end{figure}

\begin{figure} 
\centering
\includegraphics[width=7.5cm]{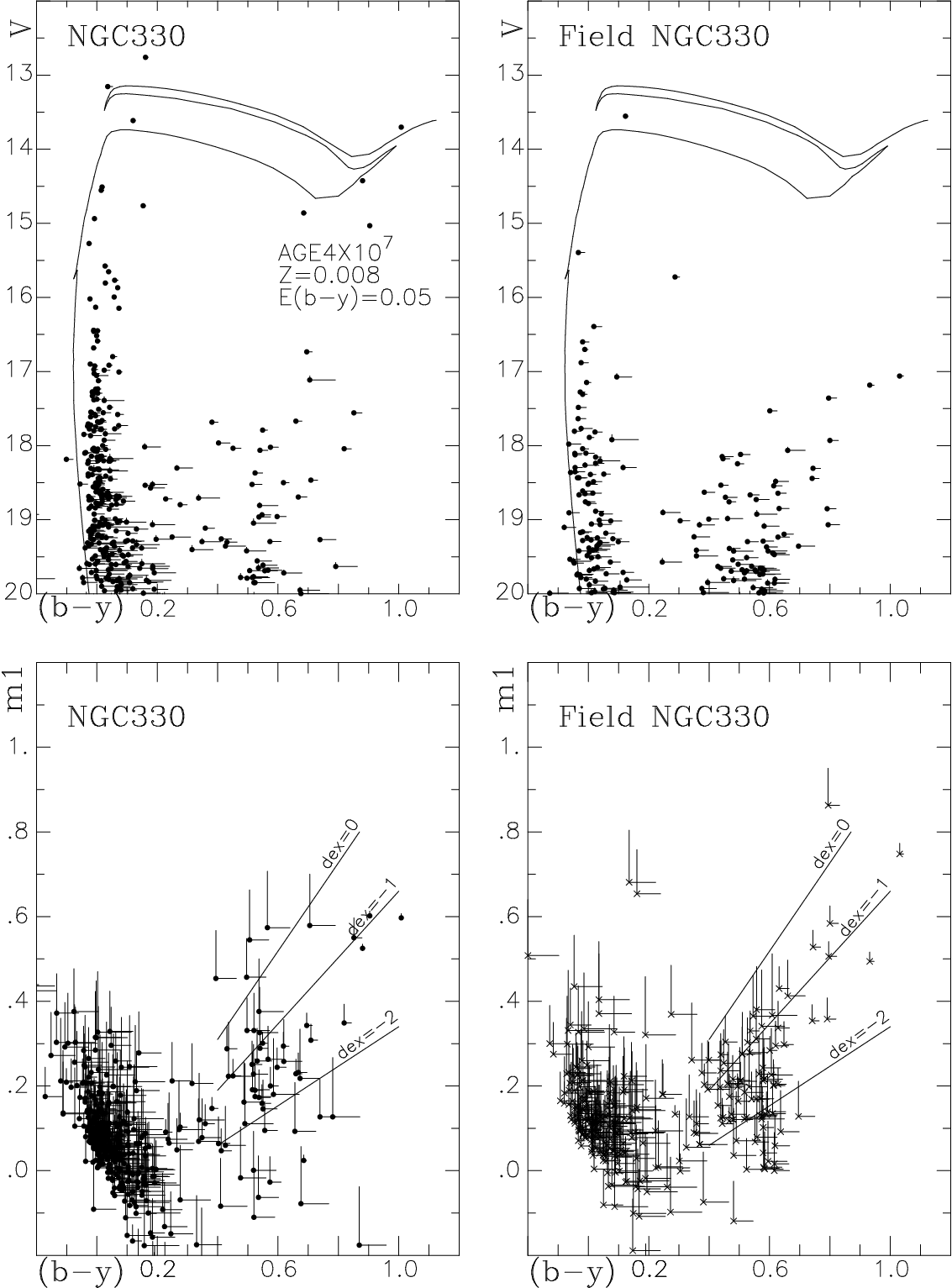}
\tiny
\caption{ Same as Fig. ~\ref{f03}  for the cluster NGC330 and its, adjoining field.}
\label{f22}
\end{figure}

\begin{figure} 
\centering
\includegraphics[width=7.5cm]{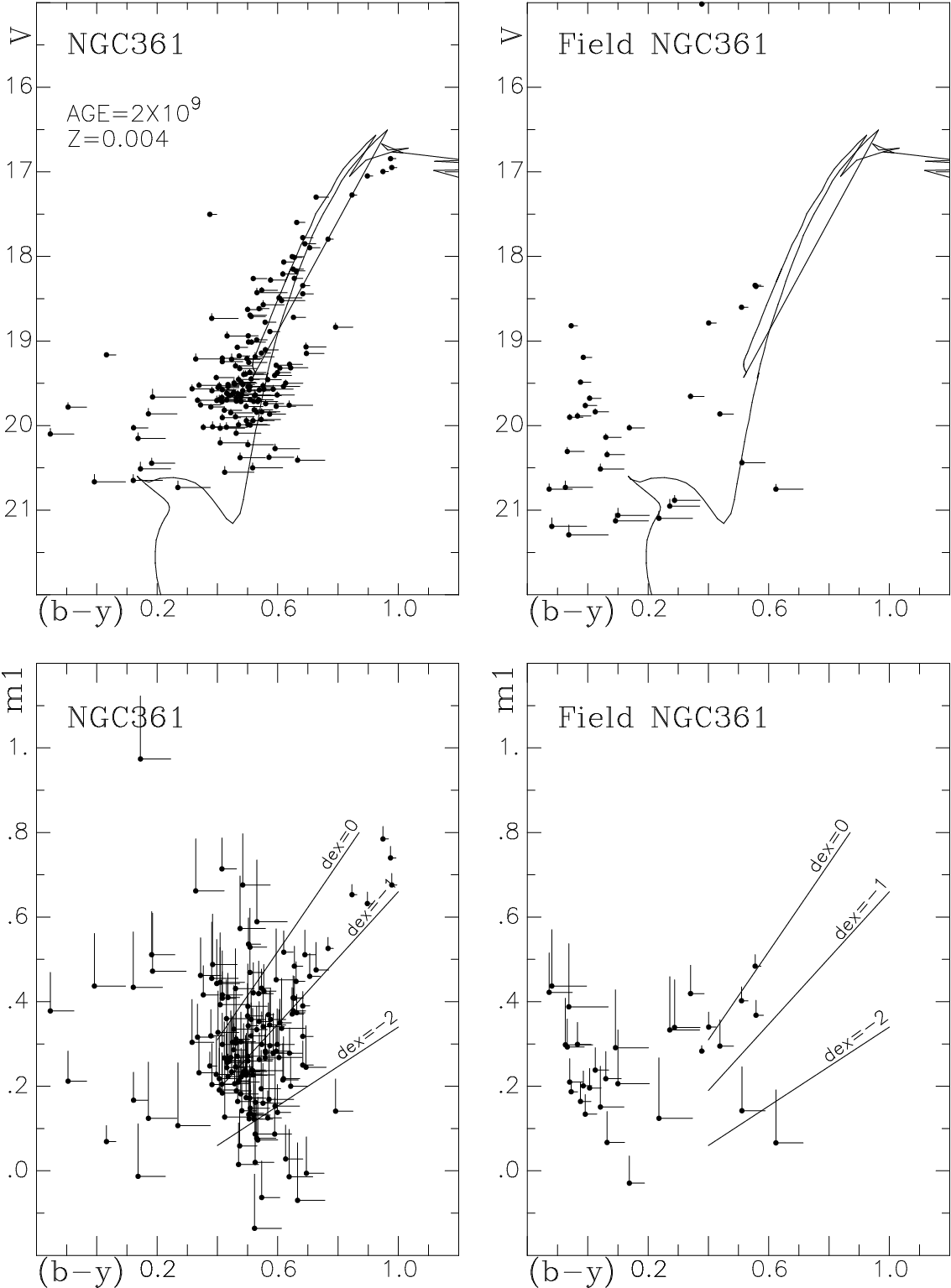}
\tiny
\caption{ Same as Fig. ~\ref{f03}  for the cluster NGC361 and its, adjoining field.}
\label{f23}
\end{figure}

\begin{figure} 
\centering
\includegraphics[width=7.5cm]{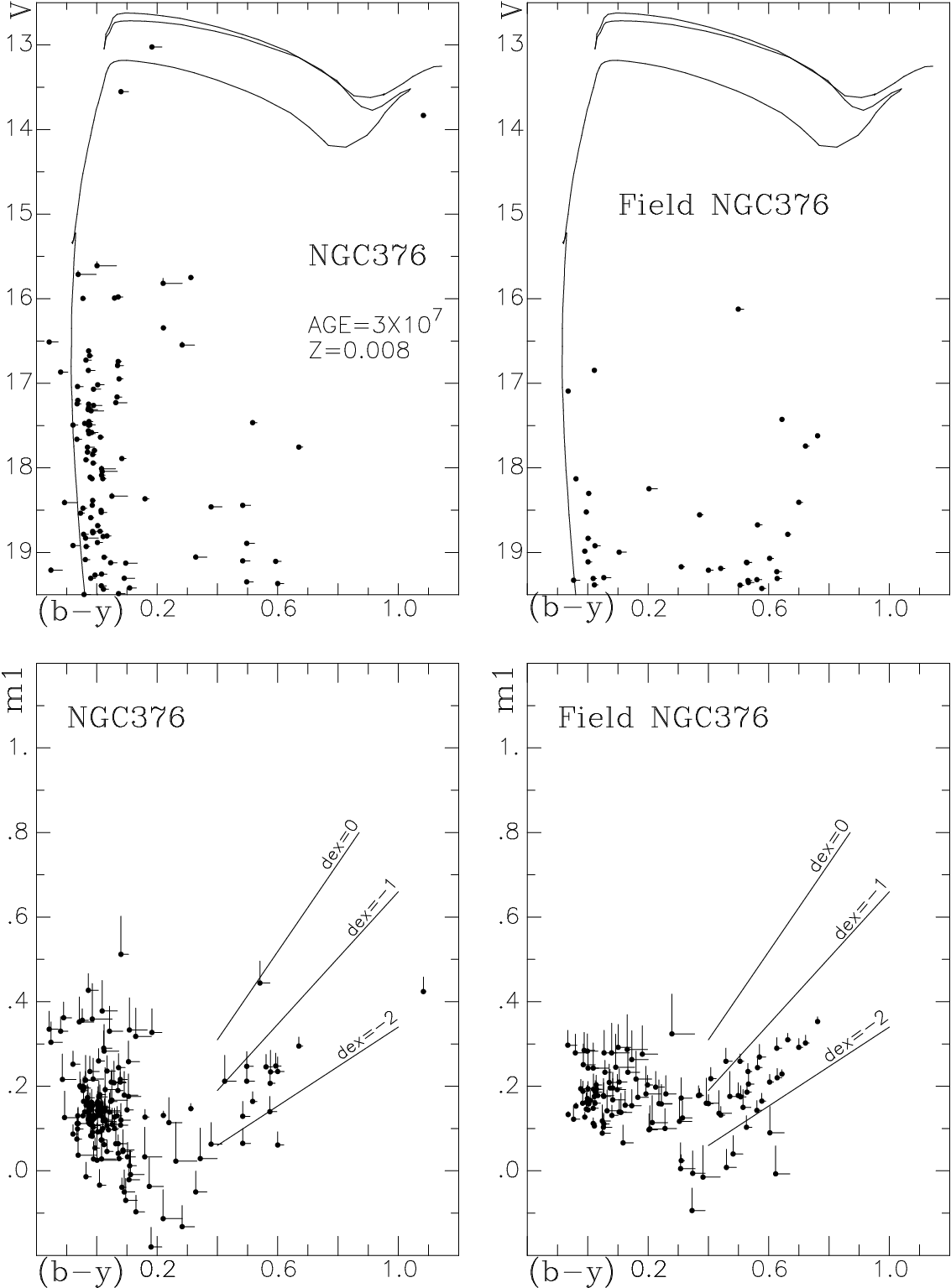}
\tiny
\caption{ Same as Fig. ~\ref{f03}  for the cluster NGC376 and its, adjoining field.}
\label{f24}
\end{figure}

A second set of diagrams for the cluster-field pairs was produced in order to derive the metallicity from the Str\"{o}mgren 
magnitudes and the traditional diagram $m_{1}$, $(b-y)$. Hilker et al. \cite{Hilker95} have produced three lines of constant 
metallicity providing the determination of the metallicity with acceptable accuracy for the late type stars, with $0.4<b-y<1.1$.
The pairs of $m_{1}$, $(b-y)$ diagrams are shown in Fig.~\ref{f03}, to Fig.~\ref{f25} (lower two diagrams) for the clusters 
and their adjoining fields respectively.
In the lower left diagram the models for the Str\"{o}mgren metallicity by Hilker et al. \cite{Hilker95} are overplotted.

Initially we have to trace the red supergiants of the cluster on the $V, (b-y)$ CMD. We again have to compare with red supergiants of the field. 
For the clusters with old ages there is a fair number of late type stars in the cluster that do not appear on the field CMDs.
Following the red supergiants on $m_{1}$, $(b-y)$ we compare with the model lines that indicate equal
metallicity. Thus we derive the metallicity value [Fe/H].
For some young clusters in our sample (KMK1, KMK3 and NGC376) the red giants if any are few
and the cluster CMD is very similar to the field CMD. So the metallicity derived from $m_{1}$, $(b-y)$ is of low accuracy.
We then adopted the metallicity derived from the isochrones.
The errors in metallicity are calculated according to Hilker et al. \cite{Hilker95} and the mean error is estimated to 0.3.
In Table~\ref{ages}, Column 2, we give the derived Str\"{o}mgren metallicity for each cluster.
The mean value of the differences between Str\"{o}mgren metallicity and the metallicity derived from the isochrones is 0.45 which is 
comparable with the adopted mean error.

\begin{table*}
\caption{Derived ages and metallicities for 15 LMC and the 8 SMC star clusters. The mean errors in age and [Fe/H] are 0.4 and 0.3 respectively.}
\begin{tabular}{lcccl} 
\hline
{\bf cluster} & {\bf Age} & {\bf [Fe/H]}                        & {\bf [Fe/H] }                          & {\bf E(b-y)} \\
\hline
              & {Gyr}      & {Str\"{o}mgren}                    & {Isochrones}                             &              \\
\hline
LMC    \\
\hline
KMK1           &   0.2         &      0.3 $^{1}$           &  0.3     & 0.05 \\
KMK3           &   0.1         &      0.0 $^{1}$           &  0.0     & 0.05 \\
KMK8           &   0.3         &     -0.5                  & -0.5     & 0.05 \\
KMK32          &   0.2         &      0.0                  &  0.15    & 0.0  \\
KMK49          &   0.4         &      0.0                  &  0.0     & 0.05 \\
KMK50          &   0.4         &      0.0                  &  0.0     & 0.05 \\
KMHK81         &   2.0         &     -1.3                  & -0.7     & 0.0  \\
KMHK1042       &   2.0         &     -1.0                  & -1.0     & 0.0  \\
KMHK1278       &   0.4         &     -0.5                  & -0.5     & 0.05 \\
KMHK1381       &   0.8         &     -1.2                  & -0.7     & 0.03  \\
KMHK1399       &   1.0         &     -1.2                  & -0.7     & 0.0  \\
KMHK1640       &   2.0         &     -1.3                  & -0.7     & 0.05 \\
HS153          &   0.2         &      0.0                  &  0.0     & 0.0  \\
SL36           &   2.0         &     -1.0                  & -0.5     & 0.0 \\
SL620          &   2.0         &     -0.5                  & -0.7     & 0.0 \\
\hline
SMC    \\
\hline 
L11            &   3.0	       &     -0.8                 & -1.3      & 0.05 \\
L17            &   3.0	       &     -1.2                 & -1.3      & 0.0  \\
L80	       &   0.2         &     -1.0                 & -0.7      & 0.0  \\
L113           &   4.0	       &     -1.7                 & -1.3      & 0.0  \\
NGC330         &   0.04	       &     -1.0                 & -0.5      & 0.05 \\
NGC361         &   2.0	       &     -0.8                 & -0.7      & 0.0  \\
NGC376         &   0.03	       &     -0.5 $^{1}$          & -0.5      & 0.04 \\
NGC419         &   1.0	       &     -1.0                 & -0.5      & 0.03 \\
\hline
\end{tabular}
\begin{tablenotes}
\small
\item 1:Adopted from the Isochrones 
\end{tablenotes}
\label{ages}
\end{table*}

Most of the clusters under investigation have not been examined before. For the  rest of them
we present values of age and [Fe/H] found in the litterature in Table \ref{litterature}. Column 1 gives the name of the cluster, columns 2, 3 list 
the age and [Fe/H] respectively. The 4th column notes the reference article. Our results are in good agreement with those found in the litterature.
Metallicity has been calculated before with Str\"{o}mgren photometry only for NGC330 (Hilker et al., \cite{Hilker95}, Grebel \& Richtler, \cite{Grebel_R_92}).
Our result (-1.0) is very close to their estimation.

\begin{table*}
\caption{Ages and metallicities for clusters found in the literature.}
\begin{tabular}{llcl} 
\hline
{\bf cluster} & {\bf Age (Gyr)} & {\bf [Fe/H]}                  & {\bf Reference} \\
\hline
\hline
L11          &  1-5              &                       & Kontizas, \cite{Kontizas_80} \\
             &  0.3 $\pm$ 0.1    &                       & Hodge, \cite{Hodge_83} \\
             &  3.5	         &   -0.80 $\pm$ 0.14    & Da Costa \& Hatzidimitriou, \cite{DaCostaHatz98} \\
             &                   &   -0.81 $\pm$ 0.13    &                       \\
L113         &  6.0 $\pm$ 1      &   -1.44 $\pm$ 0.16    & Da Costa \& Hatzidimitriou, \cite{DaCostaHatz98} \\
             &                   &   -1.17 $\pm$ 0.12    &                        \\
NGC330       &  0.007$\pm$ 0.001 &                       & Hodge, \cite{Hodge_83}   \\
             &                   &   -0.93 $\pm$ 0.16    & Hilker et al., \cite{Hilker95}, Grebel \& Richtler, \cite{Grebel_R_92}\\
NGC361       &  $>$0.5           &                       & Hodge, \cite{Hodge_83}   \\
             &  6.8 $\pm$0.5	 &   -1.45 $\pm$ 0.11    & Mighel et al., \cite{Mighel_98} \\
             &  8.1 $\pm$1.2	 &                       &                                 \\
NGC376       &  0.025 $\pm$0.01	 &       -1.08           & Piatti, \cite{Piatti07} \\
NGC419       &  0.67 $\pm$0.05   &                       & Hodge, \cite{Hodge_83} \\ 
             &  1.2-1.6          &                       & Glatt et al., \cite{Glatt_08}\\
             &  1.0-1.8          &                       & Rich, \cite{Rich_00}\\
             &  1.4 $\pm$0.2     &    -0.5 $\pm$0.25     & Piatti, \cite{Piatti11}\\
\hline
\end{tabular}
\label{litterature}
\end{table*}

\subsection{Uncertainties}

The errors in the metallicity estimations lie in three domains:
a) The photometric errors due to data reduction. 
b) Uncertainties are introduced from the selection of cluster stars, considering the contamination 
of the field stars,  and 
c) the spread of the data points around the model lines (isochrones and Str\"{o}mgren metallicity models).
However we are able to calculate arithmetically only the photometric errors and we have used on the CMDs only stars with error less or equal to $0.1mag$.
Uncertainties from the other two reasons ere visually estimated.
The uncertainties of this kind can be important in the cases of two LMC clusters: KMK1 and KMK3 and NGC376 for the SMC. 
The positions of these LMC clusters on the AMR diagram are KMK1:(0.2,0.3) and KMK3:(0.1,0.0). Thus removing them from the AMR diagram would not change either 
the trend or the discussion.
The results for KMK8, KMK1278, KMK1381 are quite satisfactory concerning the amount of uncertainty while for the rest of the clusters the uncertainty is minimum.
The mean error for all the clusters is estimated 0.4 and 0.3 for age and [Fe/H] respectively.

\subsection{Results}
After considering the previous remarks we investigate the AMR found for the fifteen clusters of the MCs.
Fig.~\ref{f26}, shows a clear trend with higher metallicities towards the youngest LMC clusters.
The accuracy of our data is within the errors described by Hilker et al. \cite{Hilker95}.

Moreover we notice a possible jump of metallicity and a considerable increase at the age of about 6$\times10^{8}yr$. This can be connected
to the latest LMC-SMC interaction which has been calculated to have happened at $10^{8}-10^{9}yr$ ago (Yoshizawa \& Noguchi \cite{YN})
The AMR for the LMC is also displaying evidence of a gradient in metallicity with distance from the centre of the cluster, since clusters 
with metallicity -1.0 to -1.5 are mainly located at the outermost regions of the galaxy Fig.~\ref{f02}.

The SMC star clusters have low metallicities regardless their location in the galaxy. No clear gradient can be found in the AMR but the 
sample is not statistically large enough to give reliable results.

\section{Conclusions}

The age-metallicity relation is a very important tool for understanding the evolution of a galaxy. 
Enhancements of metallicity may represent higher star formation activity while gaps can be associated with quiescent phases in the star formation history of a galaxy.
The age-metallicity relation (AMR) for LMC and SMC is investigated in this paper.
Moreover possible indications for gradient of metallicity in the LMC and traces of
the interaction between the two galaxies are examined. Taking into consideration the
discussion on the errors, we can summarise the results from Fig. 26 as
follows:
\begin{enumerate}
\item   The LMC displays a clear trend of AMR with higher metallicities found
in the young clusters, a result expected in a galaxy’s  normal
evolution of its stellar content. The SMC does not show such evidence,
possibly because of the small sample we have used  or because of a
different history of star formation in this galaxy.
\item    An observed jump in the LMC, shows an increase in metallicity at ages
about 6$\times$10$^{8}$. This could be the result of the most recent encounter in the
LMC-SMC that has produced an intense star formation in the LMC.
\item    A clear metallicity gradient is observed in the LMC.  The clusters with
metallicities -1.0 to -1.5 are those found in the outer regions of the
LMC.  This is an indication that the recent star formation in the LMC
occurs in the central regions.
\item    In the SMC there is no indication of an AMR relation. However this
investigation displays again the known result, that the LMC is more metal
rich than the SMC galaxy.
\end{enumerate}

\clearpage

\clearpage

\clearpage

\clearpage

\clearpage

\clearpage

\begin{figure} 
\centering
\includegraphics[width=7.5cm]{f25_NGC419.ps}
\tiny
\caption{ Same as Fig. ~\ref{f03}  for the cluster c419 and its, adjoining field.}
\label{f25}
\end{figure}

\clearpage

\begin{figure} 
\centering
\includegraphics[width=8cm]{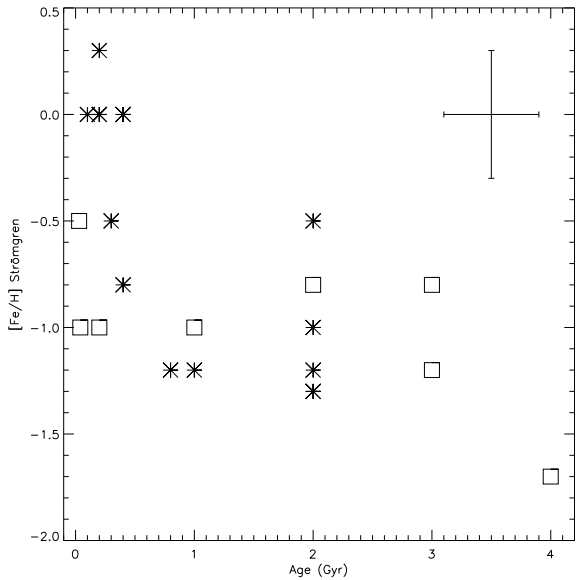}
\caption{The age-metallicity relation for LMC (asterisks) and SMC (squares) star clusters. The representative mean errors in age and [Fe/H] are 0.4 Gyr and 0.3 respectively. The corresponding error bars are ploted in the top right hand corner.}
\label{f26}
\end{figure}

\begin{acknowledgements}
The authors would like to acknowledge the NATO grant, CRG.GRGP$/$972234 and the Greek General Secretariat of Research and Technology for financial support.
B. Nordstr\"{o}m and J. Andersen acknowledge support from the Danish Natural Research council (grants 10$-$084349, 09$-$062384) and the Carlsberg Foundation.
\end{acknowledgements}

\end{document}